\documentclass[aps,10pt,prd,notitlepage,showpacs,nofootinbib,superscriptaddress, compressed]{revtex4-1}
\pdfoutput=1 
\usepackage[utf8]{inputenc}
\DeclareUnicodeCharacter{2212}{-}
\usepackage{graphicx}
\usepackage{amsmath}
\usepackage{amsfonts}
\usepackage{amssymb}
\usepackage{xcolor,soul}
\usepackage{epstopdf}
\usepackage{xcolor}
\usepackage{float}
\usepackage{subfigure}
\usepackage{hyperref}
\usepackage{rotating}
\usepackage[hyphenbreaks]{breakurl}
\usepackage[left=2cm,right=2cm,top=2cm,bottom=2cm]{geometry}
\usepackage{relsize}
\usepackage{graphics}
\usepackage{mathrsfs}
\usepackage{booktabs}
\usepackage[normalem]{ulem}
\usepackage{cancel}
\usepackage{tcolorbox}
\usepackage{tabularx,ragged2e} 
\hypersetup{
     colorlinks   = true,
     citecolor    = blue,
     linkcolor    = blue,
     urlcolor     = blue,
}
\newcommand{\bea}{\begin{eqnarray}}
\newcommand{\eea}{\end{eqnarray}}
\newcommand{\mdm}{m_\text{DM}}
\newcommand{\Edm}{E_\text{DM}}
\newcommand{\Mzp}{M_{Z'}}
\newcommand{\Erec}{E_{\rm rec}}

\newcommand{\gs}{g_\star}
\newcommand{\gss}{g_{\star s}}

\makeatletter
\begin{document}
\title{Dark matter-electron scattering and freeze-in scenarios in the light of $Z^\prime$ mediation}
\author{Basabendu Barman}
\email{basabendu.b@srmap.edu.in}
\affiliation{Department of Physics, School of Engineering and Sciences, SRM University-AP, Amaravati 522240, India}
\author{Arindam Das}
\email{adas@particle.sci.hokudai.ac.jp}
\affiliation{Institute for the Advancement of Higher Education, Hokkaido University, Sapporo 060-0817, Japan}
\affiliation{Department of Physics, Hokkaido University, Sapporo 060-0810, Japan}
\author{Sanjoy Mandal}
\email{smandal@kias.re.kr}
\affiliation{Korea Institute for Advanced Study, Seoul 02455, Korea} 
\begin{abstract}
We investigate dark matter (DM)-electron scattering in a minimal $U(1)_X$ extension of the Standard Model (SM), where the DM can appear as a Majorana fermion, a complex singlet scalar or a Dirac fermion. To study bounds on the $U(1)_X$ gauge coupling $(g_X)$ and new gauge boson mass $(M_{Z^\prime})$, from DM-electron scattering, we consider several direct search experiments like CDMS, DAMIC, SENSEI, PandaX-II, DarkSide-50 and XENON1T-S2 for different $U(1)_X$ charges. In this set-up we consider DM production via freeze-in both in radiation dominated and modified cosmological background to project sensitivities on $g_X-M_{Z^\prime}$ plane satisfying observed relic abundance. DM-electron scattering could provide comparable, or even stronger bounds than those obtained from the electron/ muon $(g-2)$, low energy scattering and intensity frontier experiments within 0.01 GeV $\lesssim M_{Z^\prime} \lesssim$ 0.1 GeV. Constrains from freeze-in could provide stronger sensitivities for $M_{Z^\prime}\gtrsim \mathcal{O}(1)$ GeV, however, these limits are comparable to those obtained from LHCb, LEP experiments for $\mathcal{O}(10)$ GeV $\lesssim M_{Z^\prime} \lesssim 150$ GeV. In future, electron-muon scattering (MUonE), proton (FASER, DUNE) and electron/positron (ILC) beam dump experiments could probe these parameters. 
\end{abstract}
\maketitle
\section{Introduction}
Two most prominent pieces of evidence of physics beyond the Standard Model (SM) are the existence of neutrino mass that oscillation experiments support~\cite{ParticleDataGroup:2020ssz} and the dark matter~(DM), whose presence is well-established from several astophysical and cosmological~\cite{Bertone:2016nfn, deSwart:2017heh,Planck:2018vyg} observations. The origin of neutrino mass can be explained in several extensions of the SM that employ the so-called seesaw mechanisms~\cite{Minkowski:1977sc,Mohapatra:1979ia,Gell-Mann:1979vob,Magg:1980ut,Schechter:1980gr,Wetterich:1981bx,Lazarides:1980nt,Mohapatra:1980yp,Foot:1988aq,Ma:1998dn,Ma:2002pf,Hambye:2003rt,Bajc:2006ia} that can also include the DM itself~\cite{Ma:2006km}, thereby addressing the two questions in one single framework. In context with DM, the weakly interacting massive particles (WIMPs) are one of the most prominent candidates~\cite{Steigman:1984ac,Jungman:1995df} (for a review, see, e.g., Refs.~\cite{Jungman:1995df, Bertone:2004pz, Feng:2010gw}). WIMPs are assumed to be in thermal equilibrium at temperatures higher than its mass. Corresponding WIMP relic density is determined by freeze-out, an epoch when the DM annihilation rate can no longer keep up with the expansion rate of the Universe. However, strong observational constraints (see, e.g., Refs.~\cite{Roszkowski:2017nbc, Arcadi:2017kky}) on the typical WIMP parameter space motivate searches beyond this paradigm. As an alternative, feebly interacting massive particles (FIMPs) have been  proposed~\cite{McDonald:2001vt,Hall:2009bx,Bernal:2017kxu}. In the early Universe, FIMPs can be generated from either the decay or annihilation of states in the visible sector. When the SM bath temperature becomes smaller than the typical mass scale of the interaction (i.e., the maximum of the DM and the mediator mass), the generation process becomes Boltzmann suppressed, giving rise to a constant comoving DM number density. Such a scenario, as opposed to freeze-out, is referred as the freeze-in~\cite{Hall:2009bx}. The FIMP paradigm requires very suppressed interaction rates between the dark and visible sectors, which can be achieved, in its infrared (IR) version, via couplings several orders of magnitude weaker than that of weak interaction coupling strength. Such feeble coupling, in general, are beyond the reach of standard collider and DM scattering experiments.  

Direct detection~\cite{Goodman:1984dc,Schumann:2019eaa} is a cornerstone of DM search experiments, serving as a laboratory method to explore particle interactions of DM in the local DM halo. The objective is to record the rare instances when DM particles collide with a target material. The sensitivity of these experiments at low DM masses is fundamentally constrained by the minimum energy needed to produce an excitation in the material. As it has been argued in~\cite{Essig:2011nj,Lin:2019uvt}, direct detection experiments that search for nuclear recoils caused by DM scattering are ineffective when it comes to DM in the sub-GeV mass range, as in that case the nuclear recoil energy is far below the lowest thresholds achieved in typical nuclear recoil experiments. Direct detection of sub-GeV dark sectors thus requires a different approach, both theoretically and experimentally. In this case, DM-electron scattering turns out to be effective as the total energy available in the scattering is significantly larger to trigger inelastic atomic processes that could lead to visible signals. Several directions have been explored (see, for example Ref.~\cite{Battaglieri:2017aum}) to understand the capability of DM-electron searches in probing or constraining physics beyond the SM.

From the perspective of particle physics model building, all these new physics models typically extend the SM particle content with additional fields, and introduce a discrete symmetry to ensure the stability of the DM over cosmological timescales. There exist a class of models, where the generation of neutrino mass and a particle DM can be simultaneously explained under the same umbrella. The general but minimal $U(1)_X$ extension~\cite{Das:2016zue} of the SM is such a framework where an SM singlet scalar, charged under $U(1)_X$ gauge group is introduced. The most intriguing aspect of this model is that incorporating three generations of SM-singlet Right-Handed Neutrinos (RHNs), as in the seesaw mechanism for generating light neutrino masses after the breaking of $U(1)_X$ and electroweak symmetries, is not merely an option but rather the simplest solution to eliminate all possible gauge anomalies. Additionally, there is a new gauge boson $Z^\prime$ which interacts differently with the left and right handed SM charged fermions manifesting a chiral scenario.   

Motivated from these, in this work we have explored the potential of DM-electron scattering experiments in constraining the $U(1)_X$ gauge coupling $(g_X)$ and the corresponding mass of the new gauge boson $(\Mzp)$, that emerges from the spontaneous breaking of $U(1)_X$\footnote{In~\cite{Figueroa:2024tmn}, DM-electron scattering has been studied in context with $L_\mu-L_\tau$ symmetry.}. We consider three benchmark models, namely, $U(1)_{\rm B-L},\,U(1)_R$, chiral $U(1)_X$ and project bounds from several DM-electron scattering experiments in $g_X-\Mzp$ plane\footnote{Phenomenological study in the context with $U(1)_{B-L}$ can be found, for example, in Refs.~\cite{Khalil:2008kp,Okada:2010wd,Seto:2020udg,Eijima:2022dec,Seto:2024lik}.}. Since the coupling involved in this scenario is typically weaker than $SU(2)_L$ gauge coupling, we also discuss DM production in all of the above cases via freeze-in. We take up three different DM candidates: Majorana, scalar and Dirac, and obtain corresponding bound in $g_X-\Mzp$ plane to produce the observed DM abundance via 2-to-2 scattering of the bath particles, mediated by $Z'$. Apart from DM genesis in the standard radiation dominated background, we also explore the impact of a generalized modified cosmological background on freeze-in yield. We summarize the status of the $g_X-\Mzp$ parameter space considering bounds from direct detection, freeze-in and several other current and future energy and intensity frontier experiments. Our analysis establishes not only the potential of DM-electron scattering experiments in constraining parameters of $U(1)_X$ gauge extension of the SM, but also the ability of laboratory and collider tests in probing modified gravity/cosmological models in future.  

The paper is organized as follows. The models are discussed in Sec.~\ref{sec:model}. In Sec.~\ref{sec:scattering-rate}, we lay out the formalism of calculating the DM-electron scattering cross-section and event rates both for atomic and semi-conductor targets. The freeze-in production of DM is elaborated for the case of radiation domination and modified cosmology in Sec.~\ref{sec:freeze-in}. In Sec.~\ref{sec:result} we summarize our main results and show the resulting bounds on $g_X-M_{Z'}$ plane coming from the freeze-in and DM-electron scattering studies. Finally we conclude the paper in Sec.~\ref{sec:concl}.
\section{Models}
\label{sec:model}
As advocated in the introduction, we consider a general minimal $U(1)_X$ extension of the SM. In $SU(3)_c\otimes SU(2)_L\otimes U(1)_Y\otimes U(1)_X$ framework, the transformation property of the SM lepton doublet $(\ell_L^i)$ is $\{1,2,-\frac{1}{2}, x_\ell=-\frac{x_H}{2}-x_\Phi\}$ and that of  the right handed electron $(e_R^i)$ is $\{1,1,-1, x_e= -x_H- x_\Phi\}$. In this set-up, the SM colored sector follow the transformation property of quark doublet $(q_L^i)$ as $\{3,2,\frac{1}{6}, x_q=\frac{1}{6}x_H+\frac{1}{3}x_\Phi\}$, right handed up type-quark $(u_R^i)$ as $\{3,1,\frac{2}{3}, x_u= \frac{2}{3}x_H+\frac{1}{3}x_\Phi\}$ and right handed down type quark $(d_R^i)$ as $\{3,1, -\frac{1}{3}, x_d= -\frac{1}{3}x_H+\frac{1}{3}x_\Phi\}$, respectively. The SM-singlet RHNs $(N_R^i)$ are transformed as $\{1,1,0,x_\nu=-x_\Phi\}$ where $i$ represents three generations of the fermions. The SM Higgs doublet and SM-singlet beyond the SM (BSM) scalar transform as $\{1,2,-\frac{1}{2}, \frac{x_H}{2}\}$ and $\{1,1,0,2x_\Phi\}$ respectively. These scalars are involved in the breaking of electroweak and $U(1)_X$ symmetries. Here $x_i$'s correspond to the $U(1)_X$ charge of the relevant SM and beyond the SM fields.
The Yukawa interactions among the scalars and fermions in the general $U(1)_X$ extension of the SM can be written as
\bea
{\cal L}^{\rm Y}= - Y_u^{\alpha \beta} \overline{q_L^\alpha} H u_R^\beta
                                - Y_d^{\alpha \beta} \overline{q_L^\alpha} \tilde{H} d_R^\beta 
				 - Y_e^{\alpha \beta} \overline{\ell_L^\alpha} \tilde{H} e_R^\beta 
			- Y_\nu^{\alpha \beta} \overline{\ell_L^\alpha} H N_R^\beta- \frac{1}{2}Y_N^\alpha \Phi \overline{(N_R^\alpha)^c} N_R^\alpha + {\rm h.c.}~
\label{LYk}   
\eea
where $H$ is the SM Higgs doublet and we write $\tilde{H} = i \tau^2 H^*$ considering $\tau^2$ as the second Pauli matrix. From the interaction Lagrangian in  Eq.~\eqref{LYk} we derive the following relations
\bea
\frac{1}{2} x_H^{}&=& - x_q + x_u \ =\  x_q - x_d=\  x_\ell - x_e=\  - x_\ell + x_\nu~; -2 x_\Phi^{}=  2 x_N. 
\label{Yuk}
\eea
Hence, applying anomaly cancellation conditions from~\cite{Das:2016zue}, and using Eq.~\eqref{Yuk} we obtain the $U(1)_X$ charges in terms of $x_H^{}$ and $x_\Phi^{}$ so that general $U(1)_X^{}$ charges of the SM fermions can be expressed as the linear combination of  $U(1)_Y$ and B$-$L charges. We find an interesting feature of this model fixing $x_\Phi^{}=1$ without the loss of generality and varying $x_H^{}$. We find, with $x_H^{}=-2$, the $U(1)_X$ charges of left handed fermions $\ell_L^i$ and $q_L^i$ reduce to zero which further converts the model into $U(1)_{\rm R}$ scenario. With $x_\Phi=1$ and vanishing $x_H$ the $U(1)_X$ charge assignment reduces into B$-$L scenario. Furthermore, taking $x_H=-1$, $-0.5$ and $1$, $U(1)_X$ charge of $e_R^i$, $u_R^i$ and $d_R^i$ turns out to be zero. As a result, the left and right handed fermions interact differently with the BSM neutral gauge boson $Z^\prime$ in this scenario. The renormalizable scalar potential of this model due to the presence of two scalar fields $H$ and $\Phi$ can be written as 
\bea
V= \ m_h^2(H^\dag H) + \lambda_H^{} (H^\dag H)^2 + m_\Phi^2 (\Phi^\dag \Phi) + 
\lambda_\Phi^{} (\Phi^\dag \Phi)^2 + \lambda^\prime (H^\dag H)(\Phi^\dag \Phi)\,,
\label{pot1x}
\eea
and for simplicity we assume that $\lambda^\prime$ to be very small. After the breaking of $U(1)_X^{}$ gauge and electroweak symmetries,  $H$ and $\Phi$ develop vacuum expectation values (VEVs) as 
\begin{align}
\langle H \rangle \ = \ \frac{1}{\sqrt{2}}\begin{pmatrix} v+h\\0 
\end{pmatrix}, \,\, \,
\langle \Phi \rangle \ =\  \frac{v_\Phi^{}+\phi}{\sqrt{2}}\,.
\label{scalar-1}
\end{align}
The electroweak scale is $v=246$ GeV at the potential minimum with $v_\Phi^{}$ being a free parameter. The breaking of general $U(1)_X^{}$ symmetry generates mass of the neutral $Z^\prime$ gauge boson as 
\begin{equation}\label{MZGL}
M_{Z^\prime}^{}=   g_X^{} \sqrt{ 4 x_\Phi^2 v_\Phi^{2}+ x_H^2 v^2}\,,
\end{equation}
which further reduces to $M_{Z^\prime} \simeq 2 g_X^{} x_\Phi ^{} v_\Phi^{}$ considering $v_\Phi^{} \gg v$ where $g_X^{}$ is the general $U(1)_X^{}$ gauge coupling. From Eq.~(\ref{LYk}) we find that the RHNs interact with $\Phi$ which generate the Majorana mass $M_N \ = \ Y_N/\sqrt{2} v_\Phi^{}$ for the RHNs after the breaking of general $U(1)_X^{}$ symmetry and the Dirac mass $m_{D}=Y_{\nu}/\sqrt{2} v$ involving $H$, $\ell_L$ and $N_R$ is generated after the electroweak symmetry breaking. These two masses induce the tiny neutrino mass through the seesaw mechanism followed by the flavor mixing. Hence we obtain the neutrino mass matrix as 
\begin{equation}
   m_\nu= \begin{pmatrix} 0&m_D^{}\\ m_D^T& M_N^{} \end{pmatrix}\,,
\label{num-1}
\end{equation}
and diagonalizing neutrino mass matrix, the light neutrino mass eigenvalue can be obtained as $-m_D^{} M_N^{-1} m_D^T$. Then the light neutrino flavor eigenstate $\nu^i_L$ can be written as $\nu^i_L \simeq U_{i\alpha} \nu_\alpha + V_{i \alpha} N_\alpha$ in terms of light $(\nu_\alpha)$ and heavy $(N_\alpha)$ neutrino mass eigenstates where $U_{i \alpha} \simeq (U_{\rm PMNS})_{i \alpha}$ where $U_{\rm PMNS}$ is the PMNS matrix and $V_{i \alpha} \simeq (m_D M_N^{-1})_{i \alpha}$ being the mixing between the light and heavy neutrinos through which RHNs interact with the SM gauge bosons. We consider $m_D/M_N \ll 1$ and we ignore it when we calculate $Z^\prime$ decay below. The interaction between $Z^\prime$ and the left $(f_L)$ and right $(f_R)$ handed fermions under general $U(1)_X$ scenario is given as
\bea
\mathcal{L}^{f} = -g_X (\bar{f}_L\gamma^\mu q_{f_{L}^{}}^{}  f_L+ \bar{f}_R\gamma^\mu q_{f_{R}^{}}^{}  f_R) Z_\mu^\prime\,.
\label{Lag1}
\eea
where $q_{f_{L}^{}}^{}$ and $q_{f_{R}^{}}^{}$ are general $U(1)_X$ charges of the SM left and right handed fermions which manifest the chiral scenario of this model. We calculate the partial decay widths of $Z^\prime$ into different SM fermions for a single generation as
\bea
\Gamma(Z^\prime \to f\bar{f})= N_C \frac{M_{Z^\prime} g_X^2}{24 \pi} \left[\left( q_{f_L}^2 + q_{f_R}^2 \right) \left(1 - \frac{{m_{f}^2}}{{M_{Z^\prime}^2}}\right)+ 6\,q_{f_L} q_{f_R} \frac{m_f^2}{M_{Z^\prime}^2} \right] \sqrt{1 - \frac{{4 m_{f}^2}}{{M_{Z^\prime}^2}}}\,,
\label{eq:zpffbar}
\eea
where $m_f$ is the SM fermion mass and $N_C^{}=1(3)$ for the SM leptons(quarks) being the color factor. The partial decay width of $Z^\prime$ into a pair of single generation light neutrinos is given by 
\begin{align}   
\label{eq:width-nunu}
    \Gamma(Z^\prime \to \nu \nu)
    =  \frac{M_{Z^\prime}^{} g_{X}^2}{24 \pi} q_{f_L^{}}^2\,,
\end{align} 
where tiny neutrino mass has been neglected and $q_{f_L^{}}^{}=-\frac{1}{2}x_H-x_\Phi$ represents general $U(1)_X$ charge of $\ell_L$. This vanishes for $x_H=-2$ and $x_\Phi=1$ and as a result there will be no tree level decay of $Z^\prime$ into a pair of neutrinos in the $U(1)_R$ scenario. In general $U(1)_X$ extended SM scenarios, $Z^\prime$ couples with RHNs as following 
\bea
\mathcal{L}_N= -\frac{1}{2}g_X q_{N_R} \overline{N} \gamma_\mu \gamma_5 N Z_{\mu}^\prime\,.
\label{neut}
\eea
We calculate the corresponding partial decay width of $Z^\prime$ into a pair of single generation of the heavy neutrino as
\begin{align}
\label{eq:width-NN}
    \Gamma(Z^\prime \to N_R^\alpha N_R^\alpha)
    = \frac{M_{Z'}^{} g_{X}^2}{24 \pi}\,q_{N_R^{}}^2\,\left( 1 - 4\frac{M_N^2}{M_{Z'}^2} \right)^{\frac{3}{2}}\,,
\end{align}
with $q_{N_R^{}}^{}(=x_\Phi=1)$ is the general $U(1)_X$ charge of the RHNs as mentioned before. In the following we discuss few possibilities of having DM candidates in the generic $U(1)_X$ model: 
\begin{itemize}
{ \item[(i)] {\bf Majorana DM}:} We consider a scenario, where a Majorana fermions could be introduced as a potential DM candidate. Among three RHNs in general $U(1)_X^{}$ scenario, one generation of the RHNs could have odd $Z_2$ parity to ensure its stability whereas the remaining generations of the RHNs could be $Z_2$-even. We consider $N^3_R$ as the potential DM candidate while neutrino mass and flavor mixing will be governed by $N^{1,2}_R$. Being charged under $U(1)_X$, DM interacts with the SM sector through $Z^\prime$ assuming a tiny mixing in the scalar sector. We write the interaction Lagrangian of the DM candidate as
\bea
-\mathcal{L}_{\rm DM}^{\rm M}= \Bigg\{ \frac{1}{2}Y_{N}^3 \Phi \overline{({N_{R}^3})^c} N^3_{R}+ h.c.\Bigg\}+ \frac{1}{2}g_X Q_{\chi} N^3 \gamma^\mu \gamma_5 N^3 Z_\mu^{\prime}\,,
\label{LDM-M1}
\eea
where the first term corresponds to the DM mass after general $U(1)_X^{}$ breaking $(M_N^3=\frac{v_\Phi}{\sqrt{2}}Y_N^3\equiv m_{\rm{DM}})$ and the second term corresponds to its interaction with $Z^\prime$. Here $Q_\chi$ corresponds to a general $U(1)_X$ charge assigned to the Majorana DM candidate.
{\item[(ii)] {\bf Scalar DM}:} We then take up another scenario, where a complex scalar $\Phi_1$ can be introduced as a potential DM candidate with the SM$\otimes U(1)_X$ charge assignment $\{1,1,0, Q_\chi\}$. $\Phi_1$ is $Z_2$-odd and the remaining particles are even under $Z_2$ parity. $Q_\chi$ is the general $U(1)_X$ charge of $\Phi_1$ because of which $\Phi_1$ interacts with other fields of the model exchanging $Z^\prime$. However, $\Phi_1$ can interact with scalar sector of the model through the following potential
\bea
V \supset \lambda_{\rm minx_1} (H^\dagger H)(\Phi_1^\ast \Phi_1)+\lambda_{\rm minx_2} (\Phi^\dagger \Phi)(\Phi_1^\ast \Phi_1)\,. 
\eea
We assume $\lambda_{\rm mix_{1,2}}$ to be very small in the line of scalar mixing considered in Eq.~(\ref{pot1x}) taken to be small. The interaction between $\Phi_1$ and $Z^\prime$ takes the following form
\bea
\mathcal{L}_{\rm DM}^{\rm s} = Q_\chi g_X Z^\prime_\mu \Big\{\Phi_1^\ast (\partial^\mu \Phi_1) - (\partial^\mu \Phi_1^\ast)  \Phi_1\Big\}+ m_{\Phi_1}^2 \Phi_1^\ast \Phi_1\,,
\label{LDM-1}
\eea
where $m_{\Phi_1}(=m_{\rm{DM}})$ is the mass of $\Phi_1$. 
{\item[(iii)] {\bf Dirac DM}:} Lastly, we consider an alternative scenario where general $U(1)_X^{}$ model can be extended with a weakly interacting Dirac fermion field $\chi_{L,R}^{}$ following the charge assignment $\{1,1,0, Q_\chi \}$. This Dirac field could be considered as a potential DM candidate with an assigned general $U(1)_X^{}$ charge $Q_\chi$ to ensure its stability. Hence the interaction Lagrangian can be written as  
\begin{align}
\mathcal{L}_{\rm DM}^{\rm D}= i \overline{\chi} \gamma^{\mu} (\partial_{\mu}+ i g_X Q_{\chi} Z^\prime_{\mu}) \chi + m_{\chi} \overline{\chi} \chi\,,
\end{align}
with $\chi= \chi_L+\chi_R$ and $m_\chi (=m_{\rm{DM}})$ is the mass of $\chi$. To ensure the stability of the DM we find that $Q_\chi \neq \{\pm 3 x_\Phi, \pm x_\Phi\}$ $=\{\pm 3,\pm1\}$ assuming $x_\Phi=1$ for $U(1)_X$. Except these charges other possibilities could be allowed. 
\end{itemize}
We mention that each DM scenario has been assumed to be present one at a time and we simply assume that kinetic mixing between $Z^\prime$ and SM neutral gauge boson $Z$ to be very small. With all the interactions and particle content at our disposal, we now move on to the discussions of quantifying DM-electron scattering event. 
\begin{figure}
\includegraphics[scale=0.3]{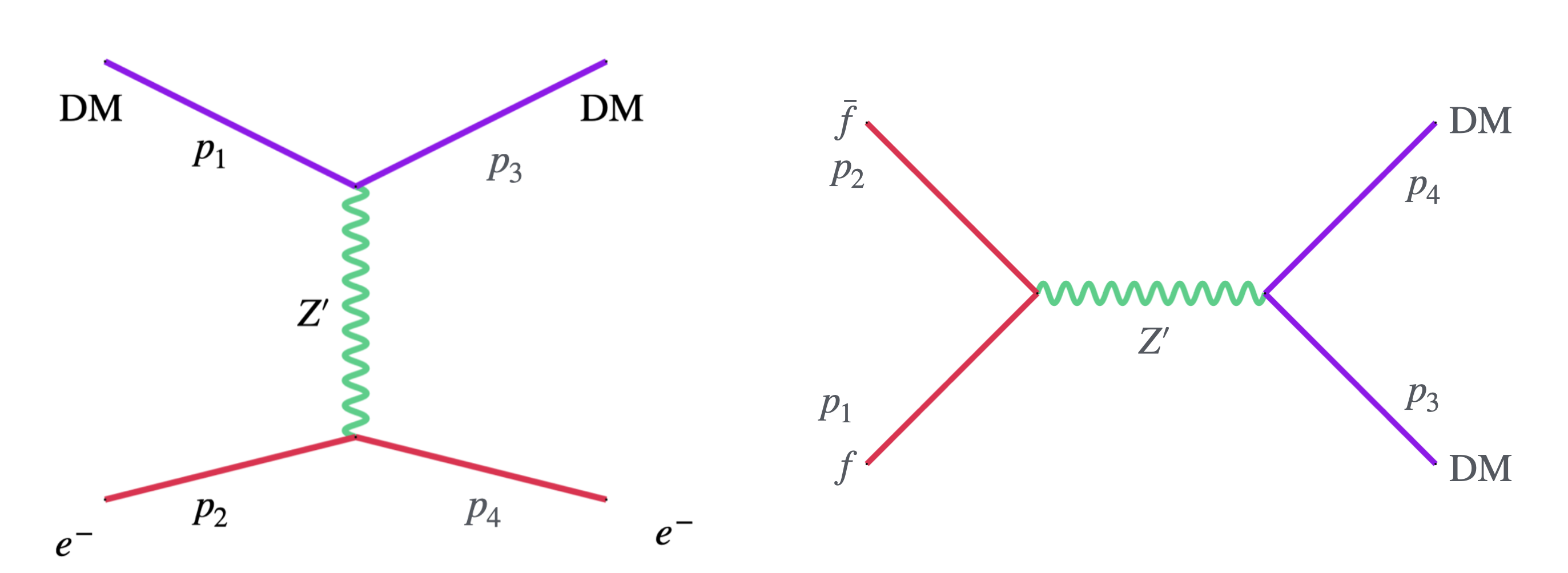}
\caption{DM-electron elastic scattering relevant for DM direct detection (left) and SM fermion annihilation into DM applicable for freeze-in (right), where DM can be a Dirac fermion, a Majorana fermion or a scalar boson.}
\label{fig:feynman}
\end{figure}
\section{Dark matter electron scattering}
\label{sec:scattering-rate}
The goal of the direct detection experiments is to observe the energy that is deposited when a galactic DM  particle scatters on the material used for the detector. The direct detection experiments are in general of two types: DM scattering on either the atomic nuclei or on the electrons. The main features of these two types of scattering are determined by kinematics. DM-atomic nuclei scattering is sensitive to DM mass scale of GeV or higher, because in order to detect the nuclear recoils, the recoil energy must be above a certain threshold and sub-GeV DM does not produce collisions with high enough energies. However, as the DM-electron scattering is {\it inelastic} in nature, it presents a viable framework for the detection of sub-GeV DM particles. This suggests that larger fraction of the DM particle's energy is transferred to the electron in the collision and ionization signals can theoretically be detected. The kinematics of the DM-electron process where the energy excites the electron to either a higher energy bound state or an ionized state is far more complex compare to DM-nuclear elastic scattering as the bound state electrons does not have definite momentum and could in fact have arbitrarily high momentum~\footnote{The scattering process may take place with any momentum transfer $q=|{\bf q}|$, however when $q$ deviates too far above the inverse Bohr radius, $a_0^{-1}\approx 3.7$ KeV, scattering rate is suppressed as it is unlikely for the atomic electron to be found with such a high momentum.}. In this case, the energy transfer to the electron is $\Delta E_e=E_{\rm rec}+E_b^i$, where $E_{\rm rec}$ is the electron recoil energy and $-E_b^i$ is the binding energy of the electron where $i$ referring to an initial state. Assuming the recoil of the atom is very small, it is possible to calculate the minimum DM velocity required for electron recoil $E_{\rm rec}$~\cite{Essig:2015cda,Radick:2020qip},
\begin{align}
v_\text{min} = {q \over 2 m_{\rm DM}} + {E_{\rm rec} + E_b^i \over q} \,.
\end{align}
Hence we see that minimum dark matter velocity required to excite an electron with recoil energy $E_{\rm rec}$ also depends on the electron binding energy.
In order to compute the recoil event rate, one needs the velocity-averaged differential cross-section, which reads~\cite{Essig:2011nj,Kopp:2009et},
\begin{equation}
\frac{d\langle\sigma_{\text{ion}}^{{i}}v\rangle}{d\ln E_{\rm rec}}=\frac{\bar{\sigma}_{e}}{8\mu^2_{e\text{DM}}}\int dq\,q|f_{\text{ion}}^{{i}}(k',q)|^{2}|F_\text{DM}(q)|^{2}\eta(v_{\rm vmin},t) \, ,
\label{eq:cross_section}
\end{equation}
where $\mu_{e\text{DM}}=m_{\rm DM}/(m_e+m_{\rm DM})$ is the reduced mass for DM-electron system, $\eta$ is the mean inverse speed and $|f_{\text{ion}}^{n\ell}(k',q)|^{2}$ is the ionization form factor. We will discuss about the mean inverse speed and ionization form factor later in this section.
The DM-electron cross-section is conventionally normalized with a reference cross-section $\bar\sigma_e$, which is defined as
\begin{align}\label{eq:ref-cs}
& \bar\sigma_e=\frac{\mu_{e\text{DM}}^2}{16\pi\,\mdm^2\,m_e^2}\,|\overline{\mathcal{M}_{e\text{DM}}(q)}|^2\Big|_{q^2=\alpha^2\,m_e^2}\,,    
\end{align}
Here $\bar\sigma_e$ is the non-relativistic (NR) DM–electron elastic scattering cross section, but with the 3-momentum transfer $q$ fixed to the reference value $\alpha\,m_e$ [cf. left panel of Fig.~\ref{fig:feynman}]. The spin-averaged squared amplitude for the matrix element encoding DM-electron scattering is given by $|\overline{\mathcal{M}_{e\text{DM}}(q)}|^2$, and are reported in Appendix.~\ref{sec:app-kin}. In reality we need to include the momentum transfer dependence and the form factor $F_{\rm DM}(q)$ in Eq.~(\ref{eq:cross_section}) captures this $q$-dependence of the matrix element: 
\begin{align}
|F_{\rm DM}(q)|^2= |\overline{\mathcal{M}_{e\text{DM}}(q)}|^2/ |\overline{\mathcal{M}_{e\text{DM}}(\alpha m_e)}|^2   .   
\end{align}
 In the NR-limit, we find the following two most interesting limits for the form factor $F_{\rm DM}(q)$, 
\begin{align}
& F_{\rm DM}(q)=\frac{\Mzp^2+\alpha^2\,m_e^2}{q^2+\Mzp^2}
\simeq
\begin{cases}
1 & M_{Z'}\gg\alpha\,m_e
\\[10pt]
\alpha^2\,m_e^2/q^2 & M_{Z'}\ll\alpha\,m_e\,.
\end{cases}
\end{align}
Since $\alpha\,m_e\simeq 3.7\times 10^{-6}$, for the parameter space of our interest, we will always have $F_{\rm DM}(q)=1$. Note that the DM-electron scattering cross-section for fixed $U(1)_X$ charges is controlled by the following set of independent parameters: $\big\{\mdm,\,\Mzp,\,g_X\big\}$.
\begin{figure*}
\centering
\includegraphics[scale=0.43]{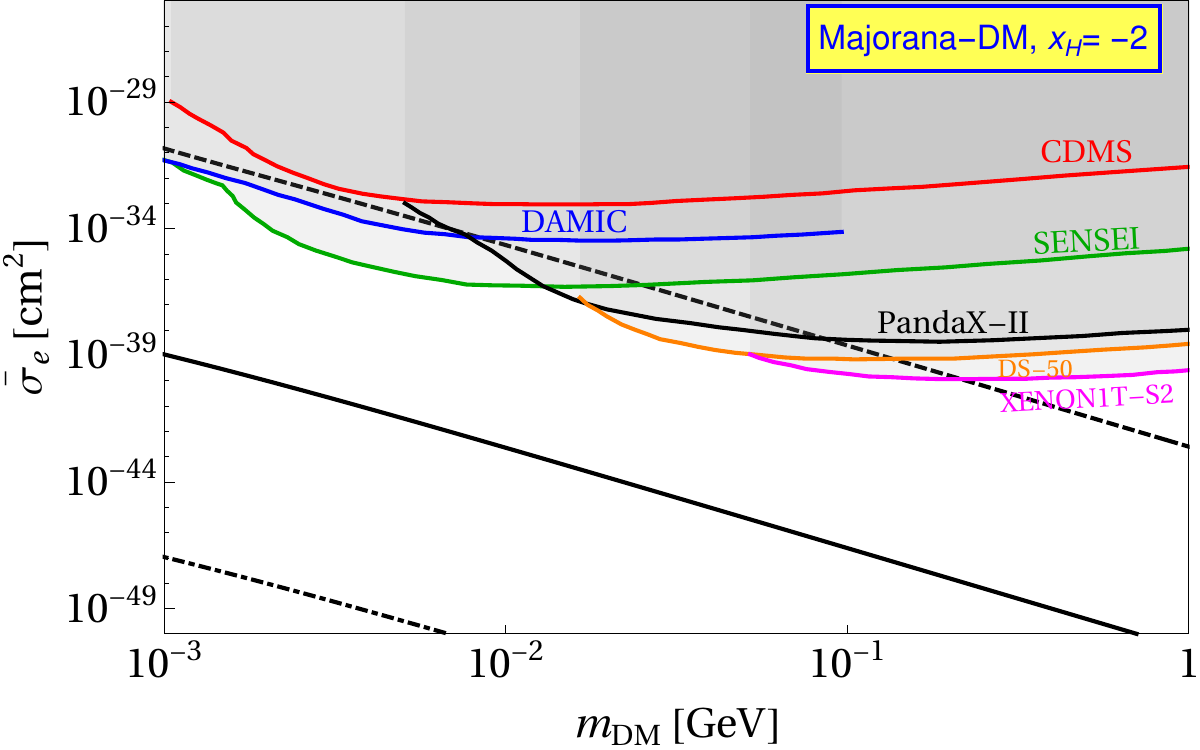}
\includegraphics[scale=0.43]{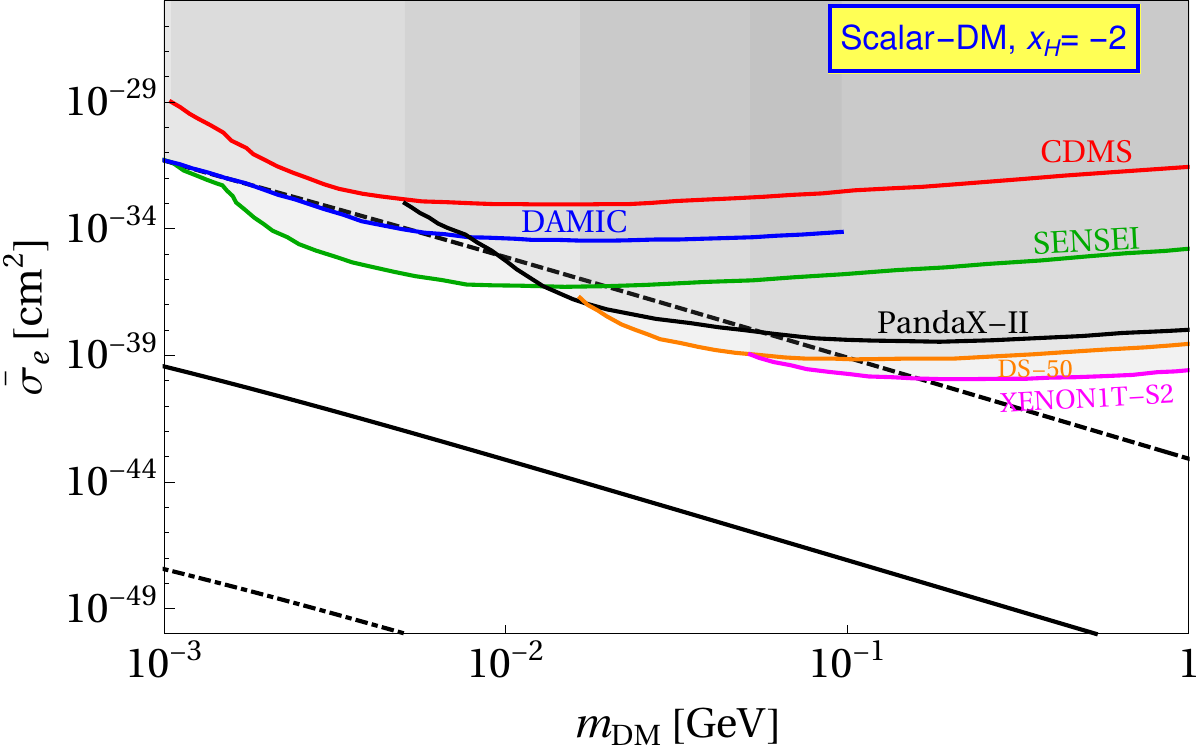}
\includegraphics[scale=0.43]{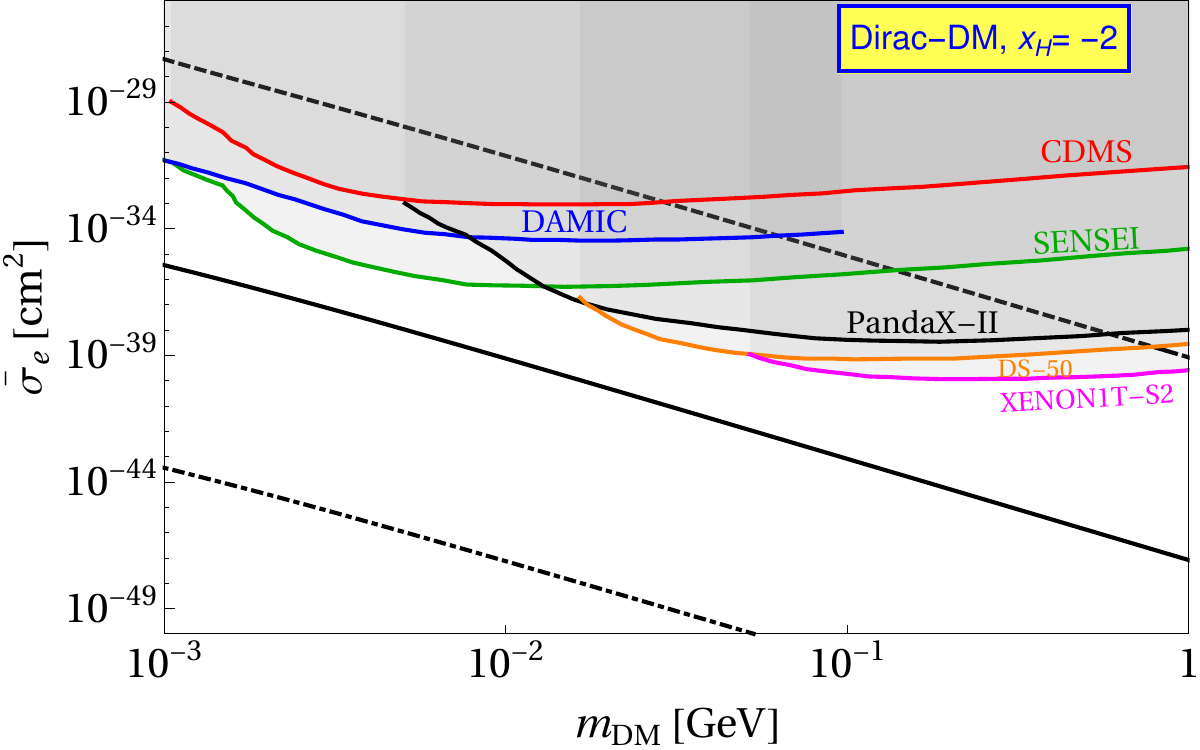}
\caption{DM-electron scattering cross-section for three benchmark points $m_{\rm DM}=M_{Z'}$~(solid black), $m_{\rm DM}=100 M_{Z'}$~(dotted black) and $m_{\rm DM}=0.01 M_{Z'}$~(dot dashed black line) with $g_X=10^{-4}$ and $Q_\chi=100$. We show existing constraints on the DM-electron scattering cross-section from Super-CDMS~\cite{SuperCDMS:2020ymb}, DAMIC~\cite{DAMIC:2019dcn}, SENSEI~\cite{SENSEI:2020dpa}, PandaX-II~\cite{PandaX-II:2021nsg}, DarkSide-50 (DS-50)~\cite{DarkSide:2022knj} and XENON1T-S2 (XE1T-S2)~\cite{XENON:2019gfn}. Here we do not impose the requirement of right DM relic density.}
\label{fig:boundDD}
\end{figure*}
In Fig.~\ref{fig:boundDD} we illustrate the DM-electron scattering cross-section for Majorana (upper left panel), Scalar (upper right panel) and Dirac DM (bottom panel), respectively for three benchmark points $m_{\rm DM}=M_{Z'}$ (solid black line), $m_{\rm DM}=100\,M_{Z'}$ (dotted black line) and $m_{\rm DM}=0.01\,M_{Z'}$ (dot dashed black line) with $g_X=10^{-4}$. We fixed the $U(1)_X$ charges as $x_H=-2$ and $x_\Phi=1$ and the charge for the Dirac DM as $Q_\chi=100$. In the same plot, we also show the existing and future constraints on the DM-electron scattering cross-section coming from experiments such as Super-CDMS~\cite{SuperCDMS:2020ymb}, DAMIC~\cite{DAMIC:2019dcn}, SENSEI~\cite{SENSEI:2020dpa}, PandaX-II~\cite{PandaX-II:2021nsg}, DarkSide-50 (DS-50)~\cite{DarkSide:2022knj} and XENON1T-S2 (XE1T-S2)~\cite{XENON:2019gfn}.
\subsection{Calculation of dark matter scattering event rate}
We see from Eq.~(\ref{eq:cross_section}) that DM-electron differential scattering rate depends on three different type of inputs: astrophysical input $\eta(v_{\rm min},t)$, atomic physics factor $|f^i_{\rm ion}(k',q)|^2$ and the particle physics input $\bar\sigma_e|F_{\rm DM}(q)|^2$. All the relevant new physics information are encoded in the particle physics input $\bar\sigma_e$ which we already discussed in the above. Here we discuss in detail about $\eta$ and $f^i_{\rm ion}$ which we need to know to compute the expected event rate in a particular direct detection experiments. The mean inverse speed $\eta$ is defined as,
\begin{equation}
\eta(v_{\rm min},t)\equiv \int_{v_{\rm min}}^{\infty}\frac{ f_\oplus \left( {\bf v}, t\right)}{v}\,d^3v,
\end{equation}
where $f_\oplus \left( {\bf v}, t\right)$ is the DM velocity distribution in the Erath frame which can be be found simply by applying a Galilean transformation to velocity distribution in Galactic frame $f_{\infty}({\bf v})$~\footnote{Assuming the velocity distribution follows the Standard Halo Model, it can be written as, 
$ f_{\infty} ({\bf v}) = 
{1 \over N_{\text{esc}} } \left( {1 \over \pi v_0^2 } \right)^{3/2} e^{- {\bf v}^2 / v_0^2 } \text{  for  }|{\bf v}| < v_{\rm esc}$ where $N_{\rm esc}$ is a normalization factor, and $v_0 \approx 220\text{ km/s}$, whereas the escape velocity is $v_{\rm esc} \approx 550\text{ km/s}$.}, so that,
\begin{align}
f_\oplus \left( {\bf v}, t\right)\approx f_{\infty} \left( {\bf v}_\odot +  {\bf v} +  {\bf V}_\oplus(t) \right) ,
\end{align}
where ${\bf v}_\odot=(11,232,7)\,\,\text{km/s}$ is the velocity of the Sun in Galactic coordinates and ${\bf V}_\oplus(t)$ is the time-dependent velocity of the Earth in the Solar frame.
The atomic physics factor $|f_{\text{ion}}^{n\ell}(k',q)|^{2}$ is the wave function suppression factor to ionize an electron in the bound state~(which is labelled as $i$) to a final state with momentum $k'$ when the momentum transfer is $q$. It is relatively easy to determine the form factor for the
atomic target compare to the semiconductor target. If the final state of the electron is a plane wave, so that $k'=\sqrt{2 m_e E_{\rm rec}}$, then for an atomic target with spherically symmetric full shell with quanta~($n,\ell$), the ionization factor is given by,
\begin{figure*}
\includegraphics[width=0.47\linewidth]{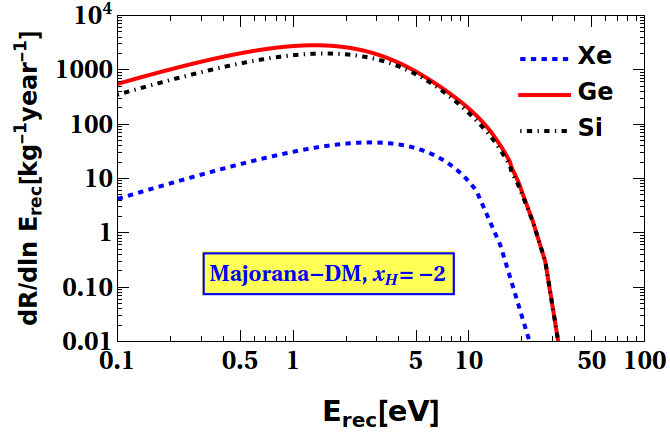}
\includegraphics[width=0.47\linewidth]{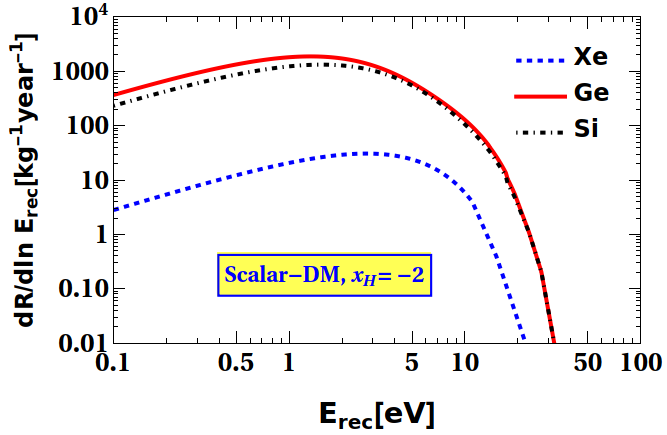}
\includegraphics[width=0.47\linewidth]{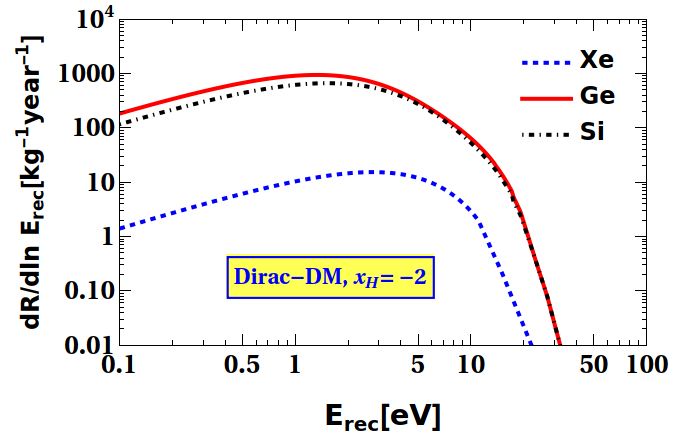}
\caption{Differential event rate for DM-electron scattering process. In each panel, the blue dashed, black dot dashed and red solid lines stands for the target Xenon, Silicon and Germanium, respectively. We fixed the DM mass as $m_{\rm DM}=10$ MeV and other relevant benchmarks for the parameters as $g_X=10^{-3}$, $m_{\rm DM}=M_{Z'}$ such that the cross-section are $\bar\sigma_e=7.3\times 10^{-40}\, \text{cm}^2$, $2.19\times 10^{-39}\,\text{cm}^2$ and $1.46\times 10^{-39}\,\text{cm}^2$ for Dirac, Majorana and Scalar DM, respectively. Note that Dirac DM cross-section is normalized by the charge $Q_\chi^2$.}
\label{fig:event_rate}
\end{figure*} 
\begin{equation}
|f_{\text{ion}}^{n\ell}(k',q)|^{2}=\frac{(2\ell+1)k'^{2}}{4\pi^3 q}\int_{|k'-q|}^{|k'+q|} dk\,k|\chi_{n\ell}(k)|^{2} \, ,
\label{atomicFF}
\end{equation}
where $\chi_{n\ell}$ is the radial part of the bound state wave function in momentum space which is given as~\cite{Kopp:2009et},
\begin{align}
\chi_{n\ell}(p)= 2\pi \int\!dr \, r^2 R_{n\ell}(r) \int\!d(\cos\theta) \, P_\ell(\cos\theta) \,e^{i p r \cos\theta},
\label{eq:chi-p}
\end{align}
where, ${\bf p}$ is a momentum space vector with modulus $p$  and $P_\ell(\cos\theta)$ is a Legendre polynomial. One can approximate the radial wave functions $R_{n\ell}(r)$ using a linear combination of orbitals known as Slater type orbitals~(STOs)~\cite{Bunge:1993jsz},
\begin{align}
R_{n\ell}(r) = \sum_j C_{n\ell j} \frac{(2Z_{\ell j})^{n_{\ell j}+1/2}}{a_0^{3/2} \sqrt{(2n_{\ell j})!}}(r/a_0)^{n_{\ell j} - 1} \exp(-Z_{\ell j} r/a_0) \,.
\label{eq:Slater}
\end{align}
Here, $a_0$ is the Bohr radius, and the parameters $C_{n\ell j}$, $n_{\ell j}$, and $Z_{\ell j}$ are taken from~\cite{Bunge:1993jsz}. Finally using Eq.~(\ref{eq:Slater}), we can evaluate Eq.~(\ref{eq:chi-p}) analytically as follows,
\small
\begin{align}
\chi_{n\ell}(p) = 
\sum_j C_{n\ell j} \, 2^{-\ell + n_{\ell j}} \bigg( \frac{2\pi a_0}{Z_{\ell j}} \bigg)^{3/2}
\bigg( \frac{i p a_0}{Z_{\ell j}} \bigg)^\ell
    \frac{(1 + n_{\ell j} + \ell)!}{\sqrt{(2n_{\ell j})!}} \,\times
    {}_2F_1 \bigg[ \tfrac{1}{2}(2 + \ell + n_{\ell j}), \tfrac{1}{2}(3 + \ell + n_{\ell j}),
\tfrac{3}{2} + \ell, -\bigg(\frac{p a_0}{Z_{\ell j}} \bigg)^2 \bigg] \,,
  \label{eq:chi-p-Slater}
\end{align}
\normalsize
with ${}_2F_1(a, b, c, x)$ being a hypergeometric function. With these above information, the differential event rate can now be obtained by summing over the all possible initial electron states as
\begin{equation}
\frac{dR}{d\ln{E_{\rm rec}}}= N_{T}\frac{\rho_{\rm DM}}{m_{\rm DM}} F(k') \sum_{i}\frac{d\langle\sigma_{ion}^{{i}}v\rangle}{d\ln E_{\rm rec}} \, ,
\label{eq:eqRate}
\end{equation}
where $N_T$ is the number of target nuclei and $\rho_{\rm DM}\approx 0.4\text{ GeV}/\text{cm}^3$  is the local DM density. $F(k')$ is the Fermi factor which accounts for how the atom itself distorts the scattered electron's wave function. In the NR limit this factor is given by
\begin{equation}
F(k')= \frac{2\pi\nu}{1-e^{-2\pi\nu}} \, \text{    with    }  \nu=Z_{\text{eff}}\,(\alpha m_{e}/k'),
\end{equation}
where $\alpha$ is the fine-structure constant and $Z_{\rm eff}$ is the effective charge that is felt by the scattered electron. Following the Ref.~\cite{Essig:2011nj}, we set $Z_{\rm eff}=1$ as this is expected to be a good approximation for outer-shell electrons. In Fig.~(\ref{fig:event_rate}), the blue dotted line shows the differential event rate, plotted against the electron recoil energy $E_{\rm rec}$  assuming Xenon atomic target. The upper left, upper right and bottom panel stands for Majorana, scalar and Dirac DM, respectively. For this plot we fixed the DM mass as $m_{\rm DM}=10$ MeV and other relevant parameters as $g_X=10^{-3}$, $m_{\rm DM}=M_{Z'}$ such that the cross-section are $\bar\sigma_e=7.3\times 10^{-40}\, \text{cm}^2$, $2.19\times 10^{-39}\,\text{cm}^2$ and $1.46\times 10^{-39}\,\text{cm}^2$ for Dirac, Majorana and scalar DM, respectively. The rate was determined using only the three outermost orbitals (5p, 5s, and 4d), which have binding energies of approximately 12, 26, and 76 eV, respectively.
\par Now let us consider the scenario where DM excites electrons in a semiconductor target above the band gap. In contrast to noble gas targets with binding energies of $\mathcal{O}(10)$ eV, semiconductor materials allow for electron ionization energies of $\mathcal{O}(1)$ eV, making them an excellent target for studying DM-electron scattering~\footnote{For example, in Germanium, electron-hole pairs are ionized to the conduction band by interactions that deposits energy above the band gap of about 0.67 eV.}. Since the electrons in a semiconductor target are characterized by Bloch wave functions in a periodic lattice, calculating the ionization form factor is particularly challenging compare to atomic target. There are many packages to calculate the ionization form factors such as EXCEED-DM~\cite{Trickle:2022fwt,Griffin:2021znd}, QEdark~\cite{Essig:2015cda} and Quantum-Espresso~\cite{Giannozzi:2009gev}. There are also analytic and semi-analytic approach to calculate the ionization form factors~\cite{Graham:2012su,Lee:2015qva}. Here, we follow Ref.~\cite{Lee:2015qva}, where form factor was derived using the Roothaan-Hartree-Fock (RHF) wave functions for the electrons. The delocalized eigenfunctions of the Hamiltonian for an electron in any periodic potential can be written in terms of localized Bloch wave functions, $\Psi_{\bf k}({\bf r})$, which may be expressed using Wannier functions~\cite{book},
\begin{align}
\Psi_{\bf k}({\bf r})=\sum_{N} e^{i\,{\bf k}.{\bf R}_N}\phi({\bf r}-{\bf R}_N),
\end{align}
where $\phi({\bf r})$ is a Wannier function localized at the site ${\bf R}_N$ and ${\bf k}$ are the wave vectors in the first Brillouin zone (BZ). The wave function of the scattered electron can be roughly represented as a plane wave for sufficiently high momentum transfers. Hence we can obtain the scattering cross section by considering the transition of an electron from a localized initial-state atomic wave function–with a ${\bf k}$-dependent binding energy–to a final-state wave function with plane-wave solution. The total differential event rate is then calculated by integrating over all binding energies and weighting it according to the density of states,
\begin{equation}
\frac{dR}{d\ln{E_{\rm rec}}}\approx N_{T}\frac{\rho_{\rm DM}}{m_{\rm DM}} F(k') \int dE_b\,\rho(E_b)\frac{d\langle\sigma_{\text{ion}}v\rangle}{d\ln E_{\rm rec}} \, ,
\label{eq:dR}
\end{equation}
where $\rho(E_b)$ is the density of states determined experimentally~\cite{ssd,Graham:2012su}, which takes into account that a given set of $E_b$ values corresponds to a different number of $\Psi_{\bf k}$ states. Hence, the density of states serves as an efficiency factor for scattering at various binding energies. In Fig.~(\ref{fig:event_rate}), the red-line (black dot-dashed line) shows the differential event rate, plotted against the electron recoil energy $E_{\rm rec}$, for DM mass $m_{\rm DM}=10$~MeV  assuming the target material is Germanium~(Silicon). In Fig.~(\ref{fig:event_rate}), when we compare the predicted differential scattering rate for the DM scattering off Germanium/Silicon with that for Xenon, we see that the rate is larger in Germanium/Silicon for all electron recoil energies; this is a direct effect of the lower binding energy in the semiconductor.
\section{Freeze-in production of dark matter}
\label{sec:freeze-in}
In this section we obtain constraints on the coupling and mass from the requirement of producing right DM abundance. Here we are typically interested in the DM-SM couplings that are feebler than weak-interaction strength, for which, as we will see, it is not possible for the DM to thermalize in the early Universe. This inevitably gives rise to DM production via freeze-in, opposed to freeze-out. In the following sections we will consider DM production takes place (i) during radiation domination (RD) and (ii) in a modified cosmological background prior to the onset of big bang nucleosynthesis (BBN). In the latter case, as we will see, the DM-SM coupling $g_X$ can be amplified by at least $\sim\mathcal{O}(10^2)$ compared to the RD scenario, depending on the choice of masses and couplings.  
\subsection{During radiation domination}
\label{sec:RD}
We first consider the scenario with standard cosmological history, where the DM genesis takes place in the background of a radiation dominated Universe. We note, the relevant DM production channels in this model are: (i) on-shell 1-to-2 decay of $Z'$ into a pair of DM final state and (ii) $Z'$-mediated $s$-channel 2-to-2 scattering of the bath particles into a pair of DM. Above the temperature of electroweak (EW) phase transition $T_{\rm EW}\simeq 160$ GeV, we consider the SM states are absolutely massless\footnote{Here we do not include temperature corrected mass terms for the particles in the bath.}, while they all become massive once the EW symmetry is broken. We therefore consider the total DM yield at present as a sum of the yield before and after the EW symmetry breaking\footnote{This treatment has been adopted, for example, in Refs.~\cite{Duch:2017khv,Barman:2020ifq,Bhattacharya:2021rwh}.}. Now, the Boltzmann equation (BEQ) governing the DM number density can be written in terms of the DM yield defined as a ratio of the DM number density to the entropy density in the visible sector, i.e.,  $Y_{\rm DM}=n_{\rm DM}/s$. The BEQ can then be expressed in terms of the reaction densities as
\begin{equation}\label{eq:beq}
x\,H\,s\,\frac{dY_{\rm DM}}{dx} = \gamma_\text{22}+\gamma_\text{12}\,,
\end{equation}
where $x\equiv\mdm/T$ is a dimensionless quantity. The complete expressions for the reaction densities $\gamma$'s due to 2-to-2 scattering $(\gamma_{22})$ and 1-to-2 decay $(\gamma_{12})$ can be found in Appendix~\ref{sec:app-reacden}. In the subsequent analysis we shall consider $Z'$ to be part of thermal bath such that its number density follows standard Maxwell-Boltzmann distribution. The Hubble parameter $H$ in the radiation dominated epoch and the entropy per comoving volume $s$ are given by
\begin{align}
& H\equiv H_{\rm rad}(T) = \frac{\pi}{3}\,\sqrt{\frac{\gs(T)}{10}}\,\frac{T^2}{M_P}\,, &  s(T) = \frac{2\,\pi^2}{45}\,\gss(T)\,T^3\,,    
\end{align}
where $T$ is the temperature of the thermal bath and $\gs(\gss)$ are the relativistic degrees of freedom (DoF) associated with the energy (entropy) density. Now, it is important to ensure that the DM does not thermalize with the SM bath in the early Universe, at least till $T=\Mzp$. For decay, this can be checked by comparing $\langle\Gamma_{Z'\to\text{DM},\text{DM}}\rangle$ with the Hubble parameter, where
\begin{align}
& \langle\Gamma_{Z'\to\text{DM},\text{DM}}\rangle= \Gamma_{Z'\to\text{DM},\text{DM}}\times\frac{K_1\left(\Mzp/T\right)}{K_2\left(\Mzp/T\right)}\,,   
\end{align}
is the thermally averaged $Z'$ decay into a pair of DM. Here $K_{1,2}$ are the modified Bessel functions and $\Gamma_{Z'\to\text{DM},\text{DM}}$ is the partial decay width of $Z'$ into a pair of DM final states. On comparison with the Hubble rate we find
\begin{align}
& \frac{\langle\Gamma_{Z'\to\text{DM},\text{DM}}\rangle}{H}\Bigg|_{T=\Mzp}\simeq\left(\frac{g_X}{10^{-8}}\right)^2\,\left(\frac{100\,\text{GeV}}{\Mzp\,\sqrt{\gs(T)}}\right)\times
\begin{cases}
7.1\times 10^{-2}\,Q_\chi^2 & \quad\text{for Dirac DM}
\\[10pt] 
3.5\times 10^{-2} & \quad\text{for Majorana DM}
\\[10pt] 
1.7\times 10^{-2} & \quad\text{for Scalar DM}\,,
\end{cases}
\end{align}
for $\mdm=\Mzp/2$. This shows, for the mass window $\Mzp$ of our interest, the out of equilibrium condition for DM production via decay can be ensured for $g_X\lesssim 10^{-8}$. 

Next, we focus on DM freeze-in through scattering and consider $\Mzp<2\,\mdm$ in order to kinematically forbid decay of $Z'$ into DM final states. Here we consider DM production from the SM bath, via $s$-channel $Z'$-mediation [cf. right panel of Fig.~\ref{fig:feynman}]. All relevant production cross-sections are reported in Appendix.~\ref{sec:app-scattering}. The condition for non-thermal production in case of scattering can be obtained by comparing the 2-to-2 scattering rate $\Gamma_{22}=\gamma_{22}/n_{\rm eq}$ with the Hubble rate, where $n_{\rm eq}=T/(2\,\pi^2)\,\mdm^2\,K_2\left(\mdm/T\right)$ is the equilibrium number density of the DM. In order to find an approximate analytical estimation, we consider $s\gg \mdm^2\,,\Mzp^2$. Using the expressions for scattering cross-sections in Appendix.~\ref{sec:app-scattering}, we obtain
\begin{align}
&\frac{\Gamma_{22}}{H}\Bigg|_{T=\Mzp}\simeq \left(\frac{g_X}{10^{-4}}\right)^4\,\left(\frac{\Mzp}{100\,\text{GeV}}\right)^2\,\frac{1}{\sqrt{\gs(T)}}
\times
\begin{cases}
1.1\times 10^{-5}\, Q_\chi^2 & \quad\text{for Dirac DM}  
\\[10pt]
5.9\times 10^{-6} & \quad\text{for Majorana DM}  
\\[10pt]
2.9\times 10^{-6} & \quad\text{for Scalar DM}\,,  
\end{cases}
\end{align}
with $\mdm=2\,\Mzp$ (such that the on-shell decay of $Z'$ into a pair of DM final states is forbidden) and $x_H=-2,\,x_\Phi=1$ (following the expressions for cross-section in Appendix.~\ref{sec:app-scattering}). Note that, this is rather a conservative estimation, since we consider equilibrium number density of the DM. Evidently, as $\sigma(s)\propto g_X^4$, it is possible to obtain the interaction rate below the Hubble rate for much larger $g_X$ compared to the decay case where the interaction rate had a $g_X^2$ dependence. As a conservative limit, we will always consider $g_X\lesssim 10^{-3}$, for freeze-in during RD. To fit the whole observed DM relic density, it is required that
\begin{equation} \label{eq:obsyield}
    Y_0\, \mdm = \Omega h^2 \, \frac{1}{s_0}\,\frac{\rho_c}{h^2} \simeq 4.3 \times 10^{-10}~\text{GeV},
\end{equation}
where $Y_0 \equiv Y_{\rm DM}(T_0)$ is the DM yield at the present epoch, that can be obtained by solving Eq.~\eqref{eq:beq}. Here $\rho_c \simeq 1.05 \times 10^{-5}\, h^2$~GeV/cm$^3$ is the critical energy density, $s_0\simeq 2.69 \times 10^3$~cm$^{-3}$ the present entropy density~\cite{ParticleDataGroup:2022pth}, and $\Omega h^2 \simeq 0.12$ the observed abundance of DM relics~\cite{Planck:2018vyg}. One can find the approximate analytical expression for the final yield as
\begin{align}
& Y_0 \simeq\left(\frac{g_X}{10^{-6}}\right)^4\,\left(\frac{100\,\text{GeV}}{\Mzp}\right)\,\frac{1}{\gss(T)\sqrt{\gs(T)}}\times
\begin{cases}
1.1\times 10^{-11}\, Q_\chi^2 & \quad\text{for Dirac DM}  
\\[10pt]
5.8\times 10^{-12} & \quad\text{for Majorana DM}
\\[10pt]
2.9\times 10^{-12} & \quad\text{for Scalar DM}\,,
\end{cases}
\end{align}
for $\mdm=2\,\Mzp$. This shows, a heavier $\Mzp$ requires larger $g_X$ in order to produce the right DM yield such that the observed DM relic abundance is satisfied. We emphasize that these are only approximate estimation for the DM yield, we however solve the full BEQ numerically to obtain the viable parameter space, taking into account the temperature-dependence of the relativistic degrees of freedom.    
\subsection{In modified cosmology}
\label{sec:mod-cosmo}
Having discussed the freeze-in mechanism during radiation domination, we now consider a scenario, where we assume the Universe prior to BBN has two different species: radiation with energy density $\rho_R$, and some other species $\varphi$ with  energy density $\rho_\varphi$. Our discussion closely follows  Ref.~\cite{DEramo:2017gpl,DEramo:2017ecx}, where this prescription has been elaborated\footnote{Phenomenological consequences of such faster-than-usual expansion scenario has been studied, for example, in Refs.~\cite{Chen:2019etb,Mahanta:2019sfo,Konar:2020vuu,Arcadi:2021doo,Barman:2021ifu,Mahanta:2022gsi,Barman:2022njh}.}. In presence of $\varphi$, total energy density of the Universe is $\rho = \rho_R + \rho_\varphi$. In case of a rapid expansion of the Universe the energy density of $\varphi$ field is assumed to be redshifted faster than the radiation. Accordingly, one can assume $\rho_\varphi\propto a(t)^{-(4+n)}$, where $a$ is the scale factor. Here $n>0$ implies $\varphi$ energy density dominates over radiation during early enough times. A general form of $\rho_\varphi$ can then be obtained using the entropy conservation $g_\star\left(T\right)^{1/3}aT=\text{constant}$ in a comoving frame as
\begin{equation}
 \rho_\varphi(T) =  \rho_\varphi(T_R)\left(\frac{g_{*s}(T)}{g_{*s}(T_R)}\right)^{(4+n)/3}\left(\frac{T}{T_R}\right)^{(4+n)}\,,
\end{equation}
where the temperature $T_R$ is an unknown variable and can be considered as the point of equality where $\rho_\varphi(T_R)=\rho_{\text{rad}}(T_R)$ is achieved. Using this, the total energy density at any temperature $T$ can be expressed as
\begin{align}\label{eq:totalrho}
\rho(T) &= \rho_R(T)+\rho_{\varphi}(T)=\rho_R(T)\left[1+\frac{g_* (T_R)}{g_* (T)}\left(\frac{g_{*s}(T)}{g_{*s}(T_R)}\right)^{(4+n)/3}\left(\frac{T}{T_R}\right)^n\right]\,.
\end{align}
The Hubble rate then reads
\begin{align}
& H(T)= H_{\rm rad}\left(T\right)\Biggl[1+\Biggl(\frac{T}{T_R}\Biggr)^n\Biggr]^{1/2} \simeq 
\begin{cases}
H_{\rm rad}(T)\,, & T\ll T_R
\\[10pt]
H_{\rm rad}(T)\,\left(\frac{T}{T_R}\right)^{n/2}\,, & T \gg T_R\,,  
\end{cases}
\label{eq:mod-hubl}      
\end{align}
where $H_{\rm rad}\left(T\right)$ is the Hubble parameter in the standard radiation dominated Universe. It is important to note from Eq.~\eqref{eq:mod-hubl} that the expansion rate is larger than what it is supposed to be in the standard cosmological background for $T>T_R$ and $n>0$. The temperature $T_R\gtrsim \left(15.4\right)^{1/n}~\text{MeV}$~\cite{DEramo:2017ecx,DEramo:2017gpl}, such that the generation of light nuclei during BBN is not hampered. For $n=3$, the lower bound on $T_R$ turns out to be 2.5 MeV. In the rest of the analysis we will fix $T_R=5$ MeV (which is just above the lower bound on BBN temperature $T_{\rm BBN}\simeq4$ MeV) and consider $n\leq 3$. Note that, this bound ceases to exist for $n=0$, as in that case we get back standard cosmology, without any new extra species driving the expansion of the Universe. As it has been explained in~\cite{DEramo:2017gpl}, within the modified cosmological setup, the DM is always under-produced with respect to the case of a standard history (RD). This can be physically understood from the fact that as DM production takes place during the epoch of fast expansion, hence the DM number density easily gets diluted, resulting in under abundance. Thus, the key point here is: in a faster-than-standard expanding universe, freeze-in production is suppressed, implying, to produce enough DM to match observations, larger coupling is required\footnote{The same inference also applies to freeze-in production of DM during an early matter dominated (EMD) era~\cite{Hardy:2018bph,Bernal:2018kcw,Cosme:2020mck}.}. This in turn improves the observational aspect for freeze-in, since now the DM-SM coupling has a larger strength compared to freeze-in production in a radiation dominated Universe. 
\begin{figure*}[htb]
\includegraphics[scale=0.43]{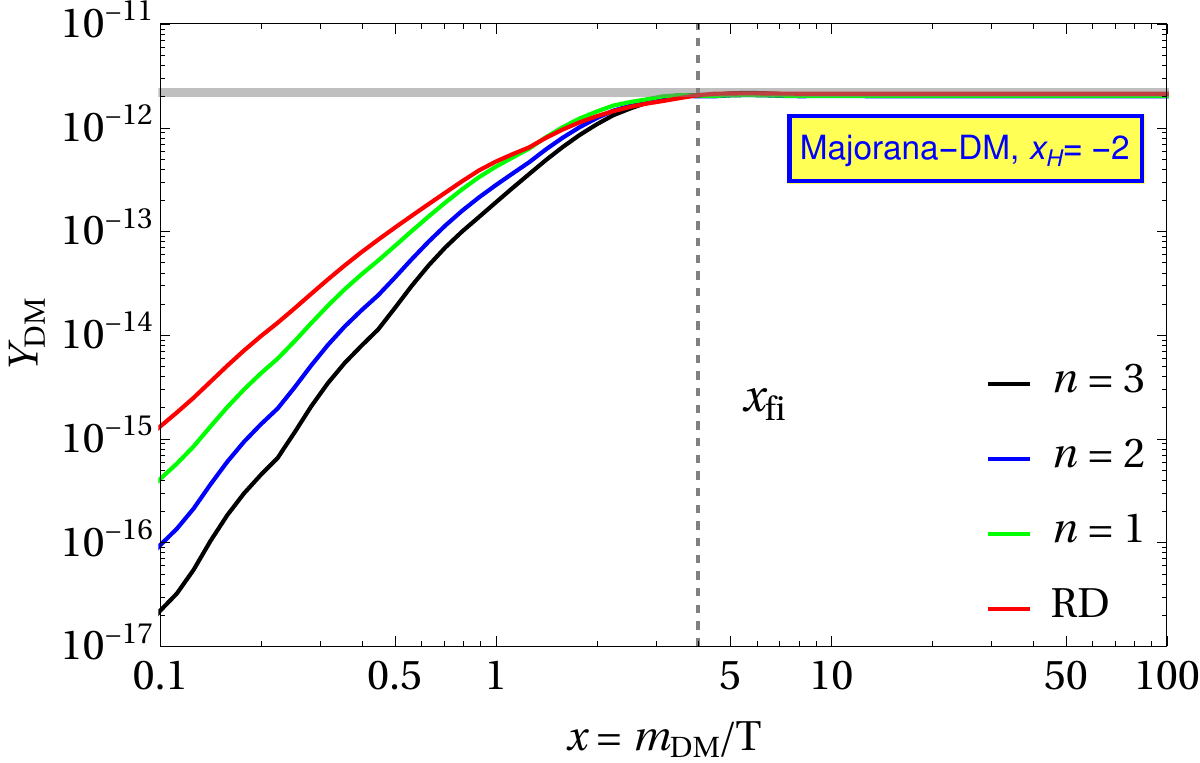}
\includegraphics[scale=0.43]{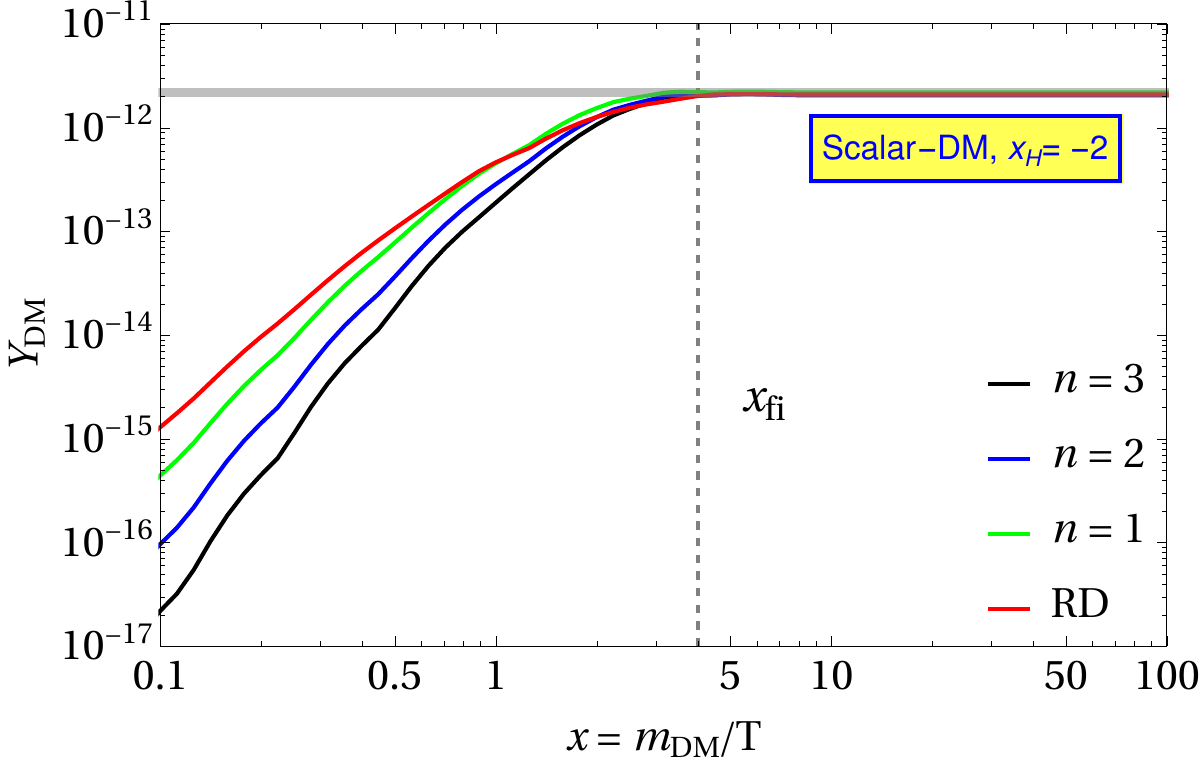}
\includegraphics[scale=0.43]{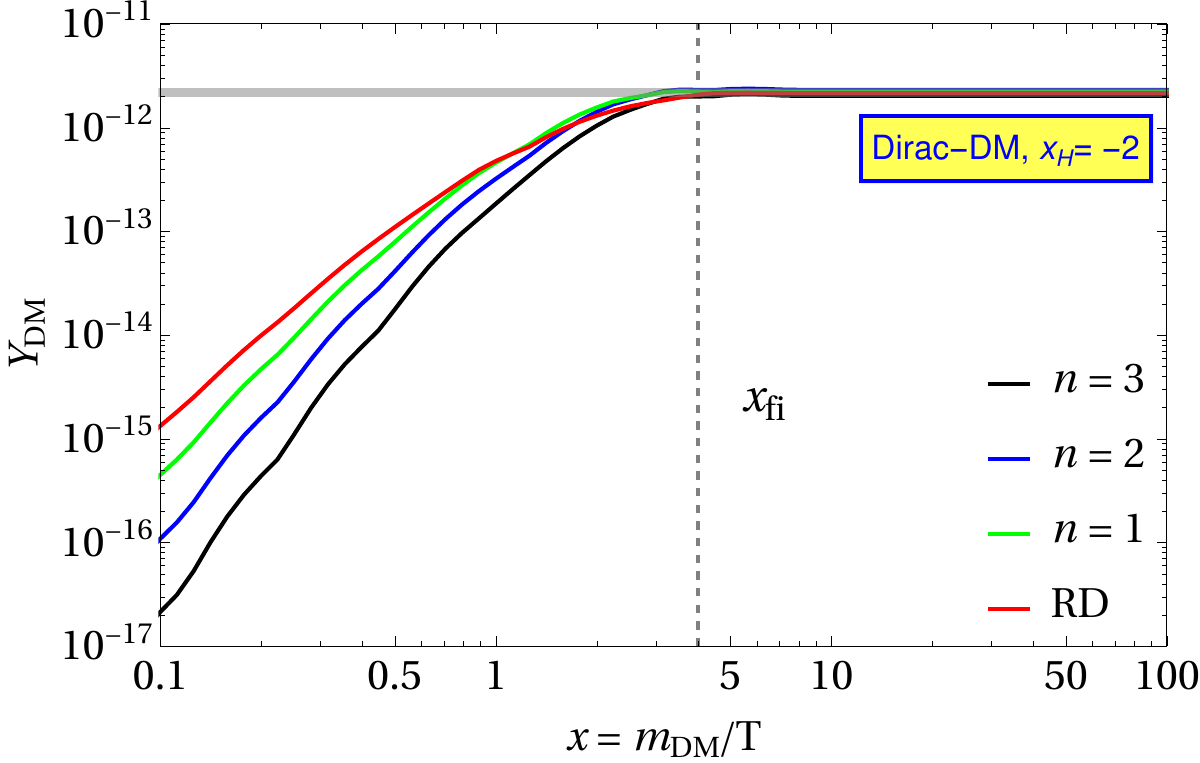}
\caption{DM yield $(Y_{\rm DM})$ as a function of $\mdm/T$, for standard cosmology (red) and modified scenario (in green, blue and black) for freeze-in production via $Z'$-mediated $\rm{SM~SM} \to \rm{DM~DM}$ scattering process. We consider $Q_\chi=100$ and $T_R=5$ MeV, with $\Mzp=\mdm/2=100$ GeV. The gray horizontal band corresponds to observed DM abundance and the vertical dashed black line, $x_{\rm fi}$, denotes the epoch of freeze-in.}
\label{fig:yield}
\end{figure*}

This is shown in Fig.~\ref{fig:yield}, where we illustrate the evolution of DM yield as a function of $\mdm/T$, for different DM candidates and for a fixed $T_R=5$ MeV. In each case the red curve corresponds to DM production in standard radiation dominated Universe. As we see, increasing $n$ results in suppression of DM production. Thus, in order to match the right relic abundance, one needs to increase $g_X$. The behaviour of the DM yield also clearly shows the IR nature of freeze-in, where freeze-in typically happens at low temperature, which in this case is given by $x_{\rm fi}\equiv\mdm/T_{\rm fi}\simeq 4$. It is necessary to mention that although the evolution of yield seems to be identical in all three cases, that is because we have fixed the DM mass, which in turn fixes the asymptotic DM yield $Y_0\simeq 2.2\times 10^{-12}\,\left(200\,\text{GeV}/\mdm\right)$, however the coupling required to produce the observed abundance for different DM spins is largely different. For instance, in case of Dirac DM, right relic abundance is produced with $g_X=\{4.8\times 10^{-6},\,1.7\times 10^{-5},\,6.0\times 10^{-5},\,2.0\times 10^{-4}\}$, corresponding to $n=\{0,\,1,\,2,\,3\}$ respectively. For the same set of masses, in case of Majorana DM, the coupling required to obtain the right abundance turns out to be $g_X=\{3.8\times 10^{-4},\,1.3\times 10^{-3},\,4.6\times 10^{-3},\,1.6\times 10^{-2}\}$. Due to faster than usual expansion of the Universe, it is easier to keep the DM out of equilibrium. Consequently, with larger $n$, it is possible to achieve larger couplings compared to the standard RD scenario, satisfying freeze-in requirements.   

Although we remain agnostic about the possible cosmological models that can incorporate such modification to the standard lore, we point towards a few well-motivated scenarios, where such modification to the Hubble expansion rate can be realized. For instance, theories with $n=2$ are quintessence fluids motivated by the accelerated expansion of the Universe~\cite{Caldwell:1997ii, Sahni:1999gb}. One possible realization of scalar potential giving rise to this behaviour is $V(\varphi)\sim \exp\left(-\lambda\,\varphi\right)$~\cite{Ratra:1987rm, Copeland:1997et}. For theories with $n>2$ one has to consider scenarios faster than quintessence. Example of such theories can be found, for example, in~\cite{Buchbinder:2007ad}, where one assumes the presence of a pre-big bang ``ekpyrotic" phase. Other than these versions of the non-standard cosmology, there is also the scope for modification of Hubble rate due to modified gravity theories~\cite{Dunsby:1997fyr,Catena:2009tm,Dent:2009bv,Leon:2013qh,Baules:2019zwk}. 

\section{Results and discussions}
\label{sec:result}
\begin{figure}
\centering
\includegraphics[scale=0.203]{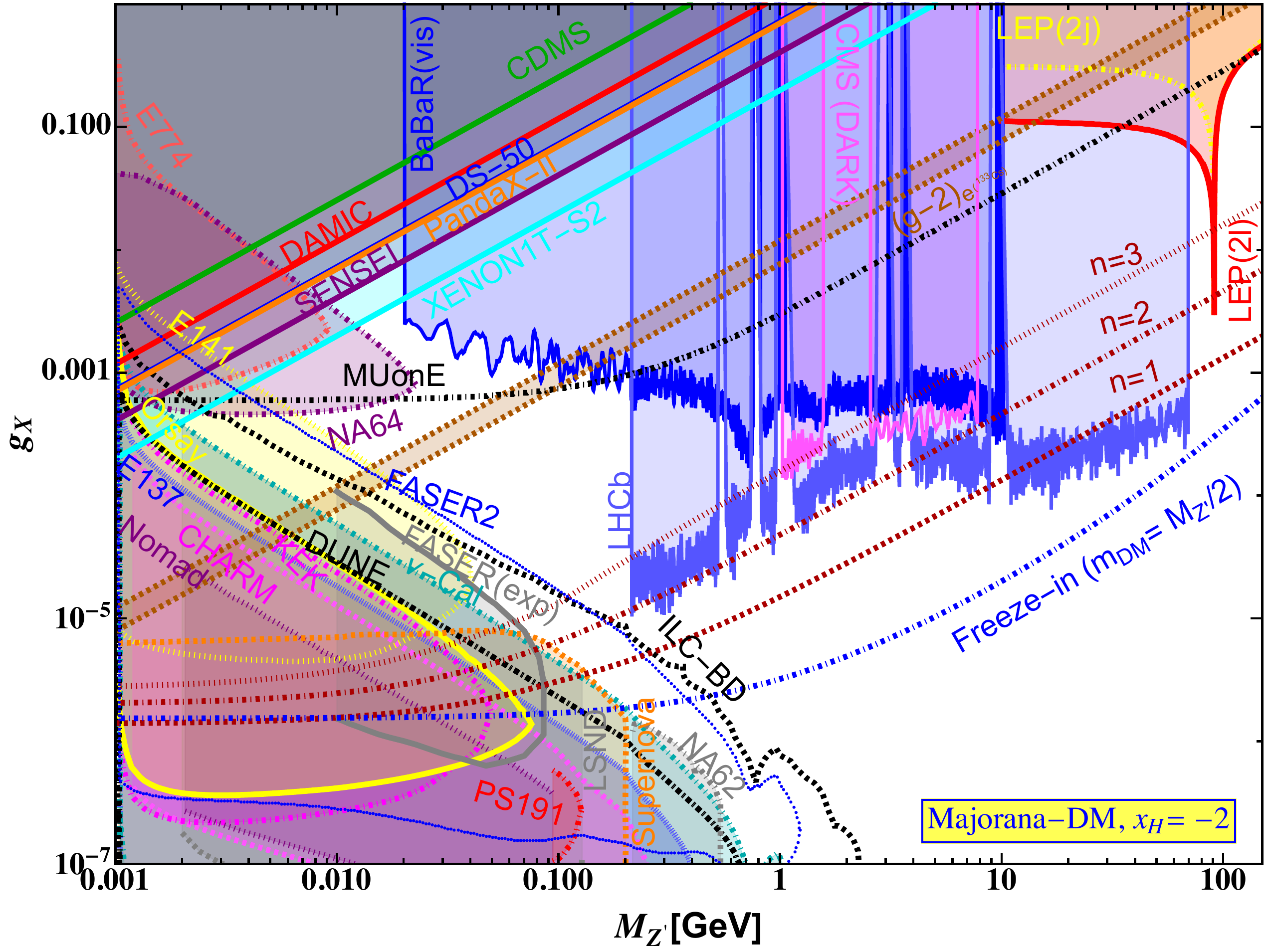}
\includegraphics[scale=0.203]{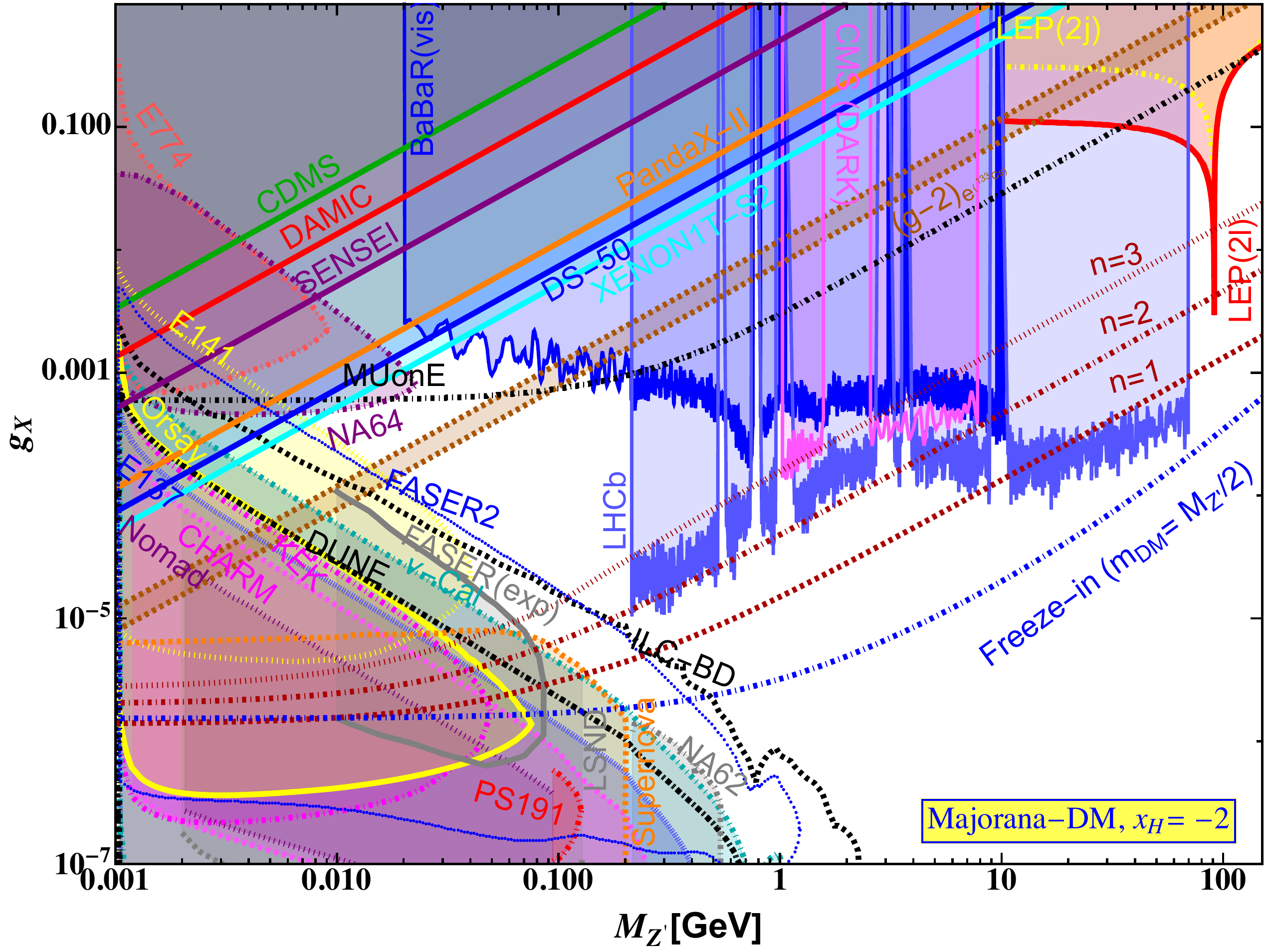}
\includegraphics[scale=0.203]{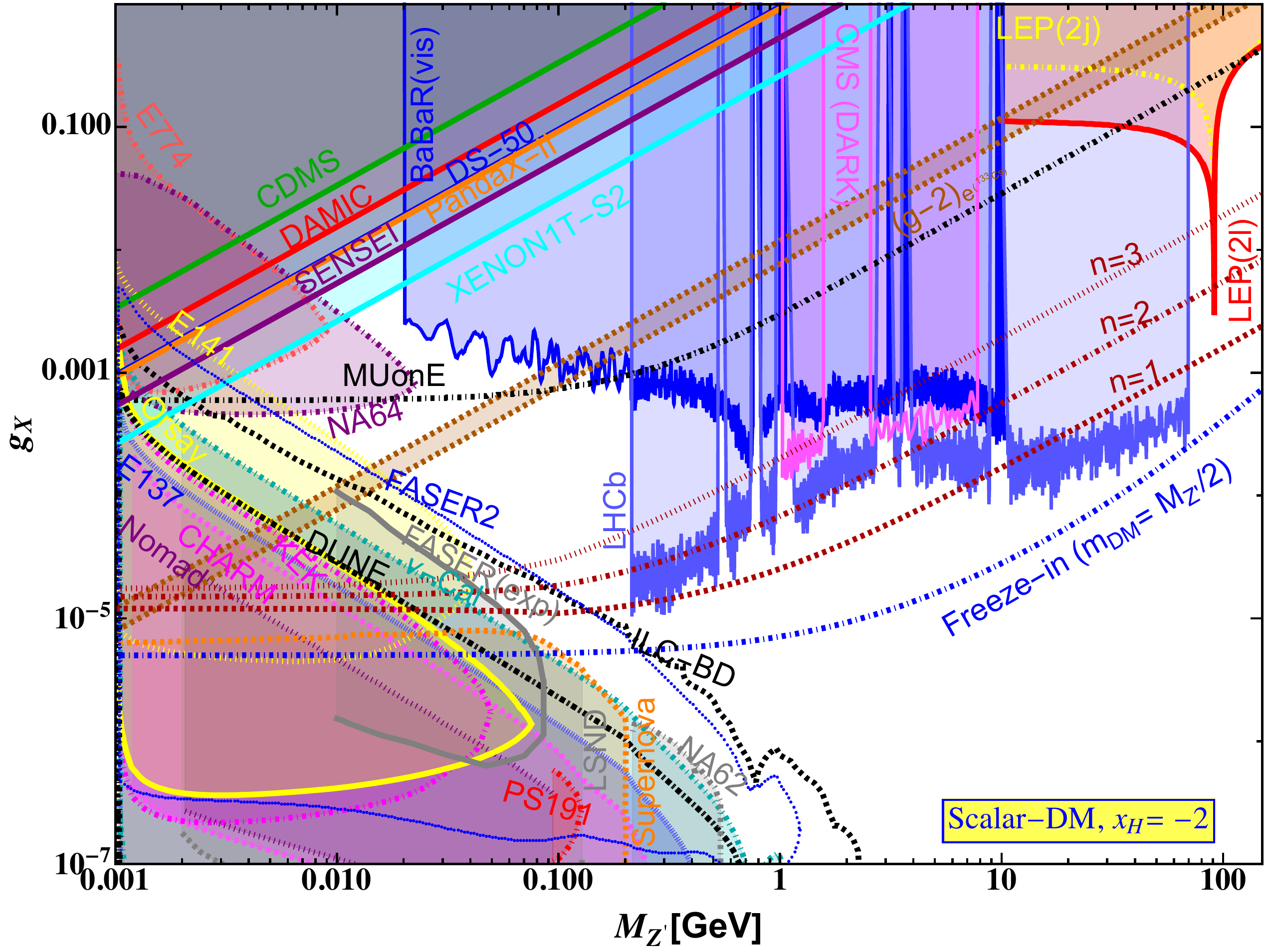}
\includegraphics[scale=0.203]{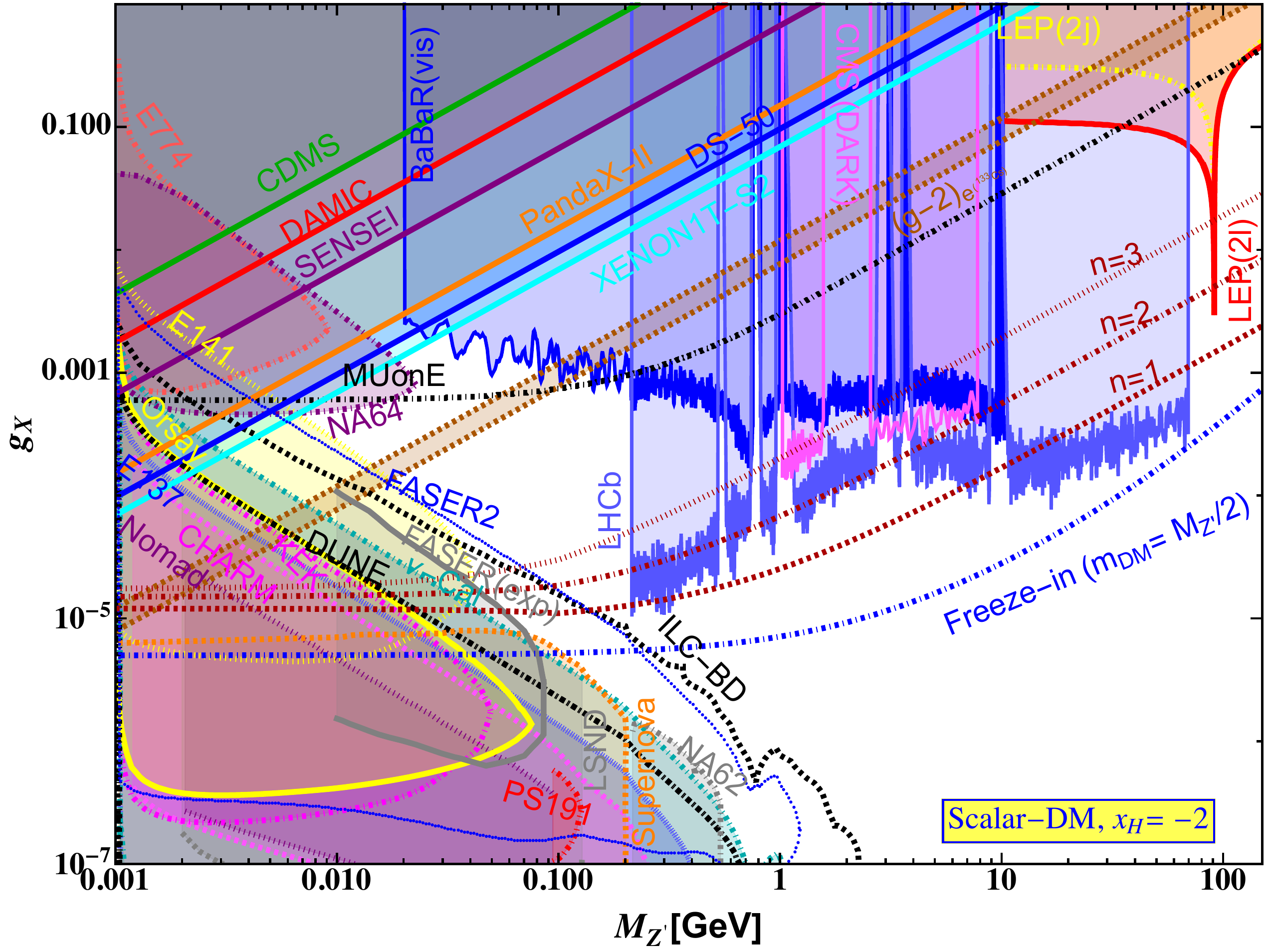}
\includegraphics[scale=0.203]{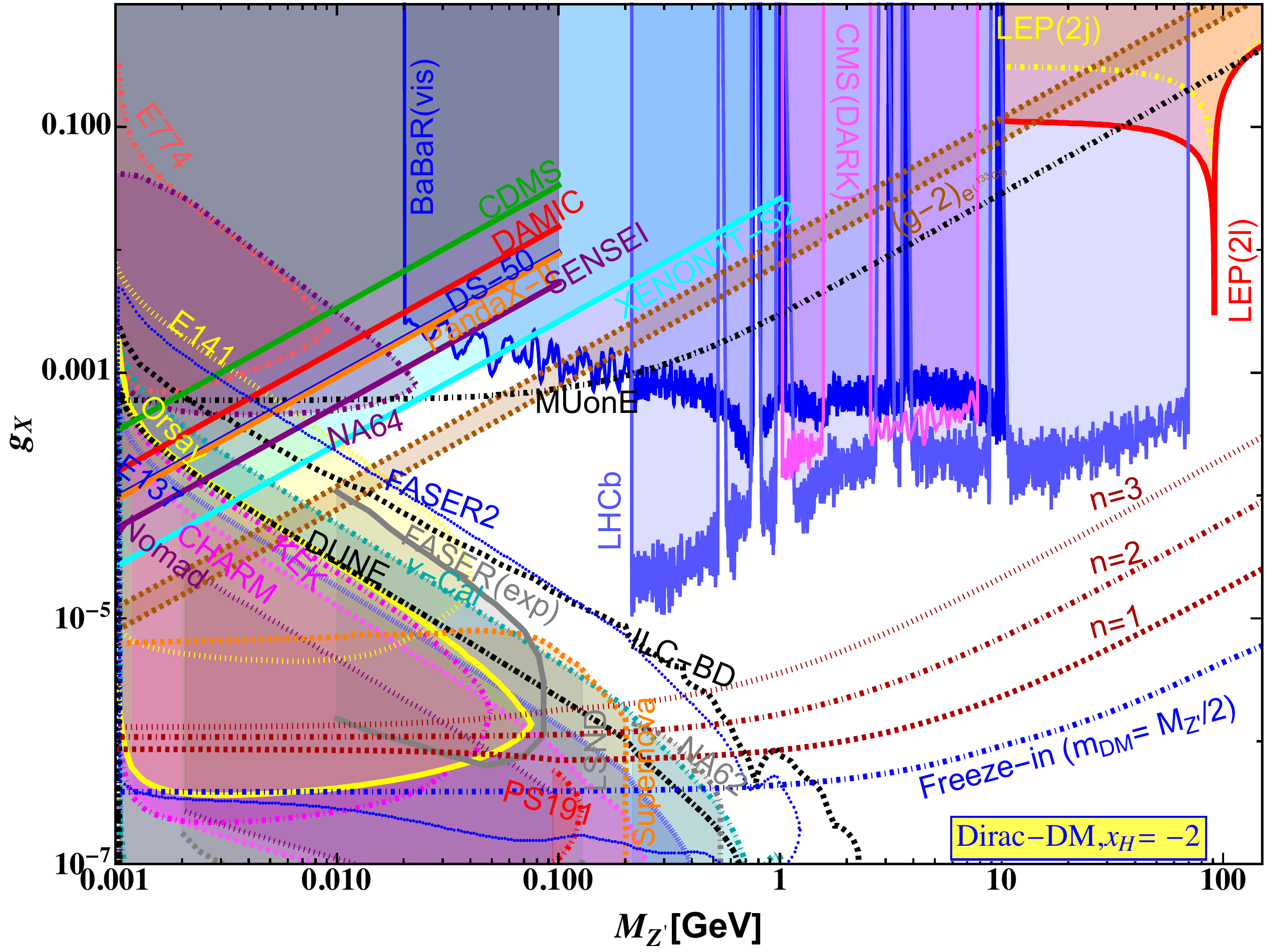}
\includegraphics[scale=0.203]{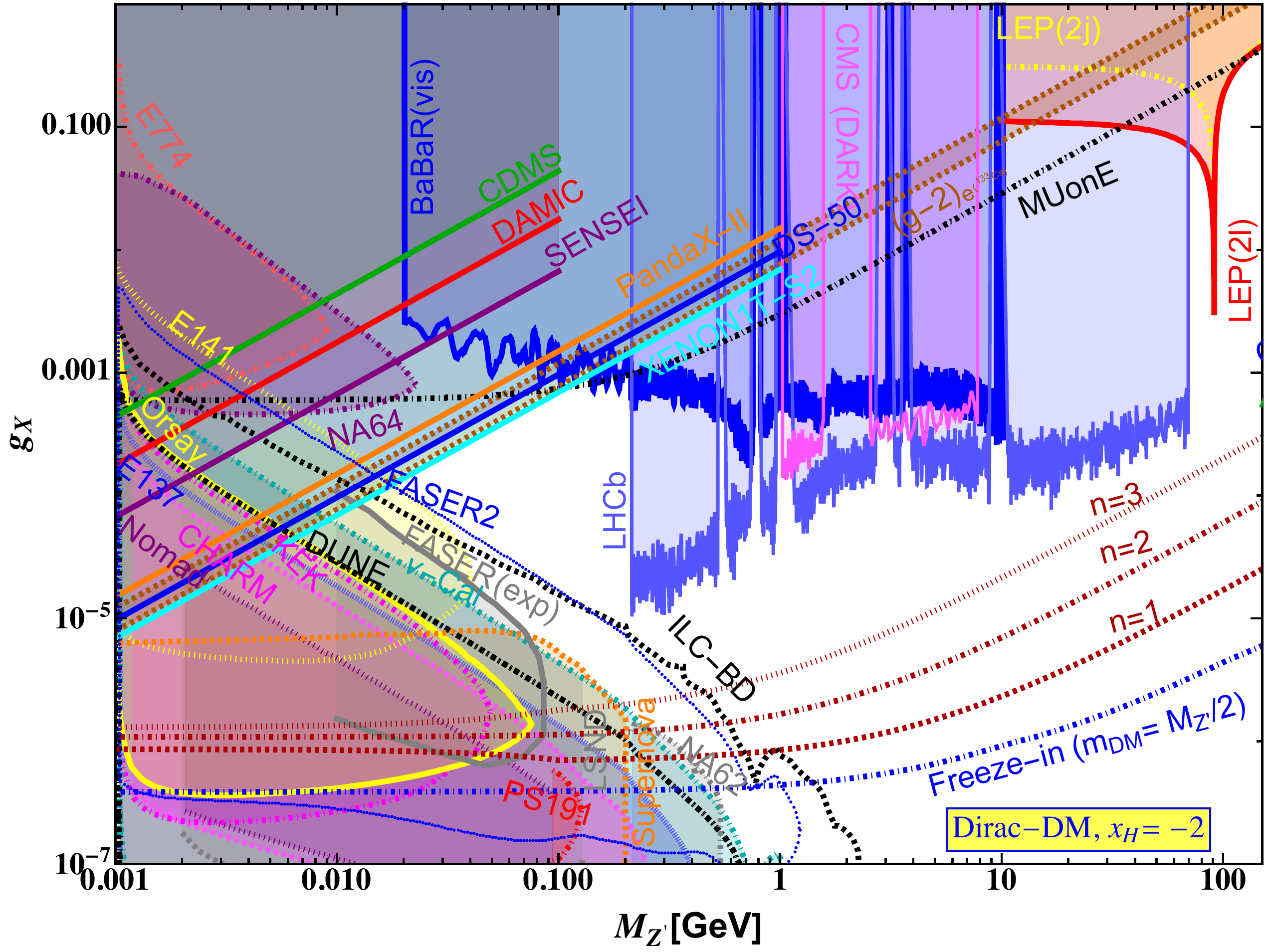}
\caption{Limits on $g_X-M_{Z^\prime}$ plane from DM-electron scattering process in $t$-channel mediated by $Z^\prime$ for $x_H=-2$ considering $m_{\rm DM}=10$(100) MeV in left (right) panel for Majorana (upper), Scalar (middle) and Dirac (lower panel) DM candidates. Shaded regions are ruled out by existing experimental data whereas prospective bounds could appear from MUonE, DUNE, FASER2 and ILC-BD experiments. We add constraints obtained from Freeze-in mechanism considering $m_{\rm DM}=M_{Z^\prime}/2$ including modified cosmology.}
\label{lim1}
\end{figure}
\begin{figure}
\centering
\includegraphics[scale=0.214]{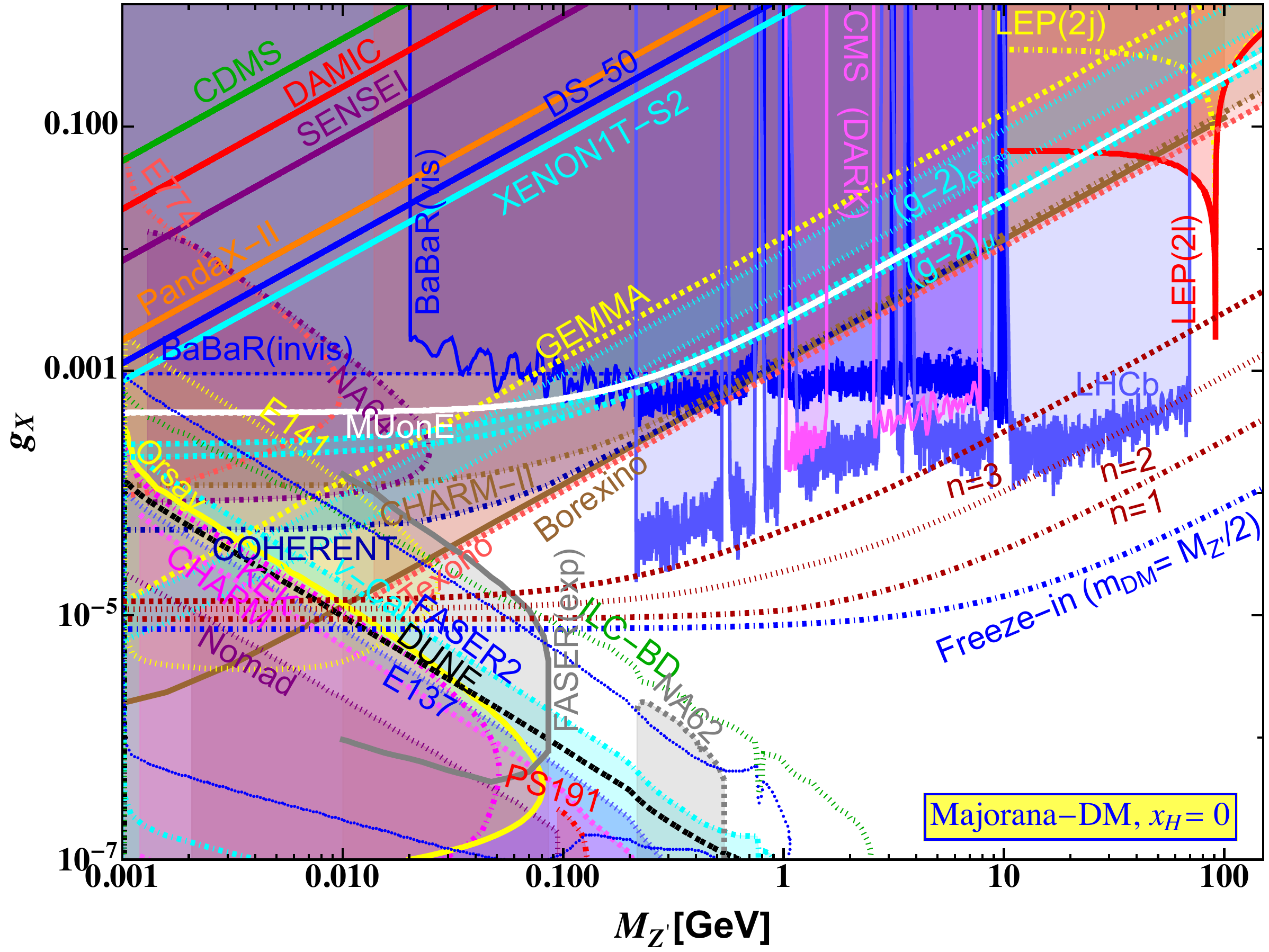}
\includegraphics[scale=0.203]{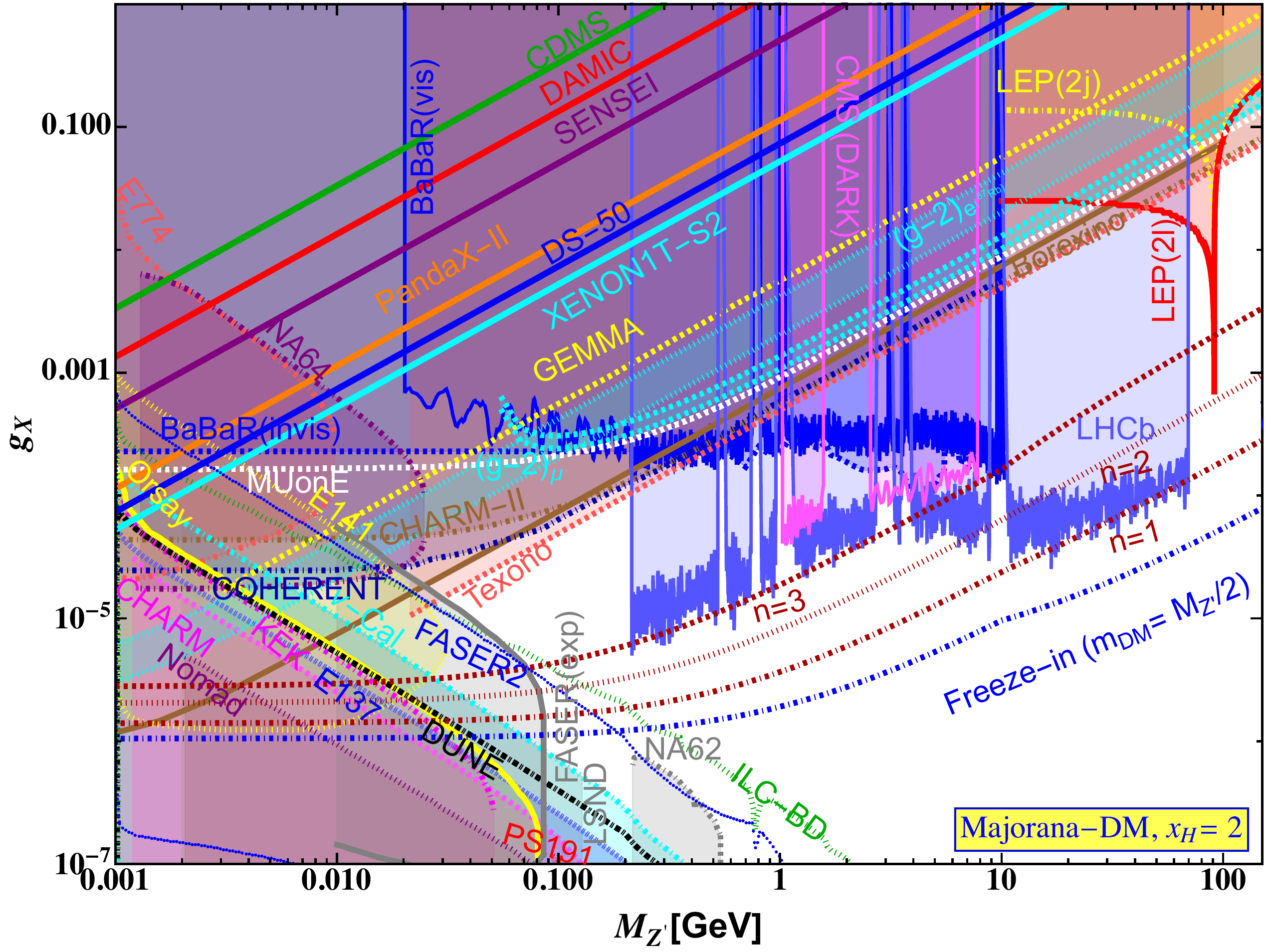}
\includegraphics[scale=0.214]{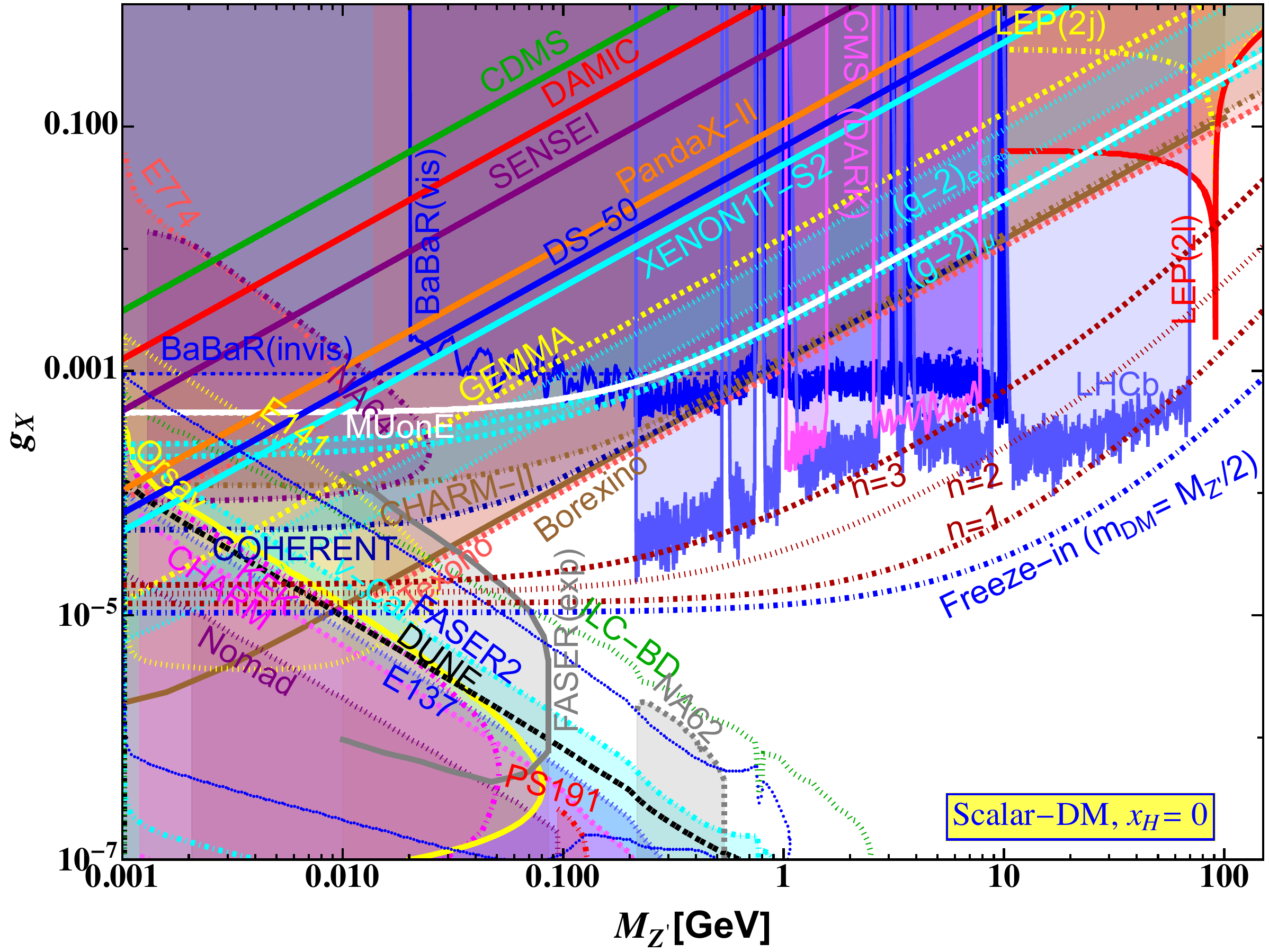}
\includegraphics[scale=0.203]{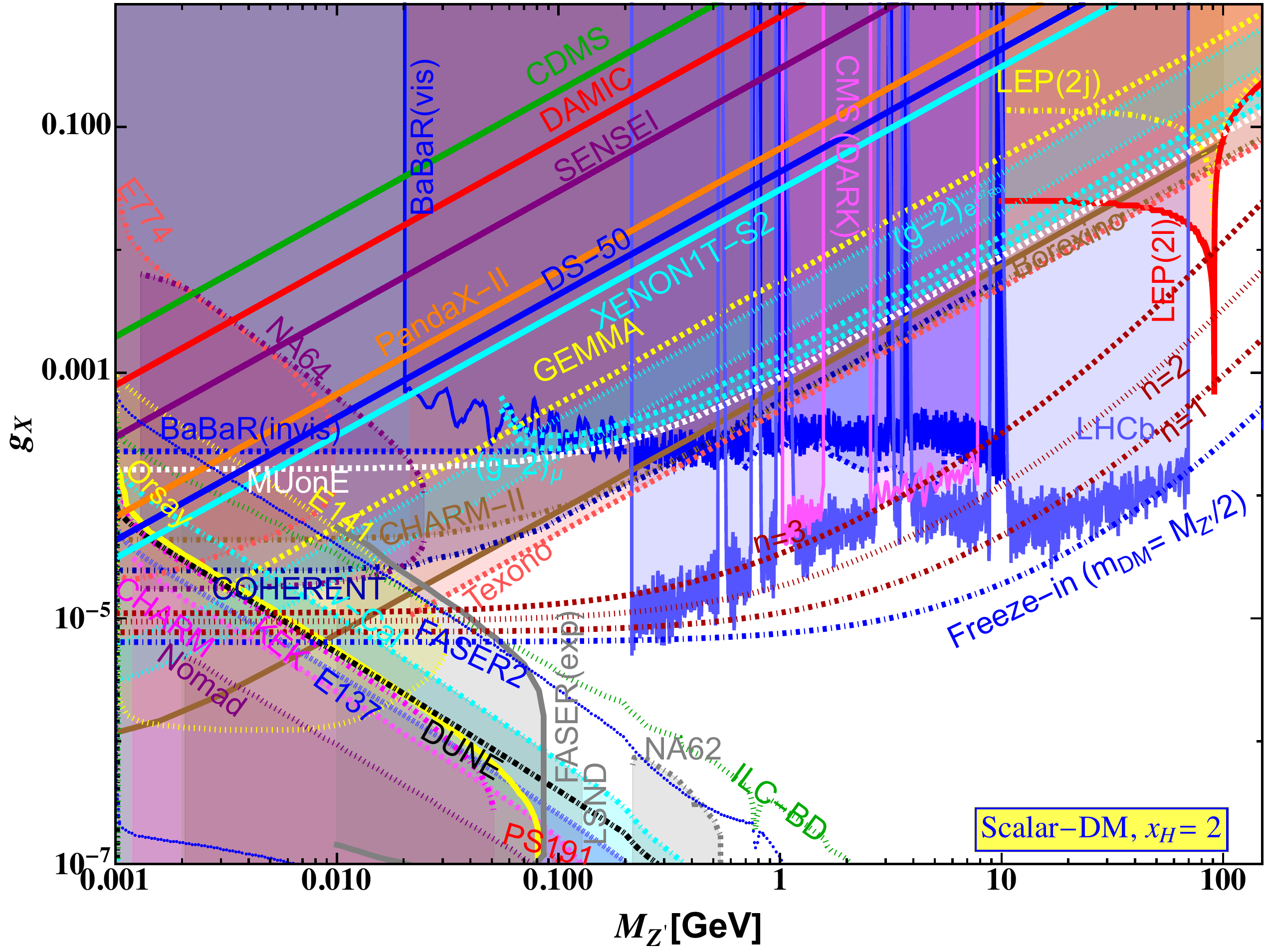}
\includegraphics[scale=0.214]{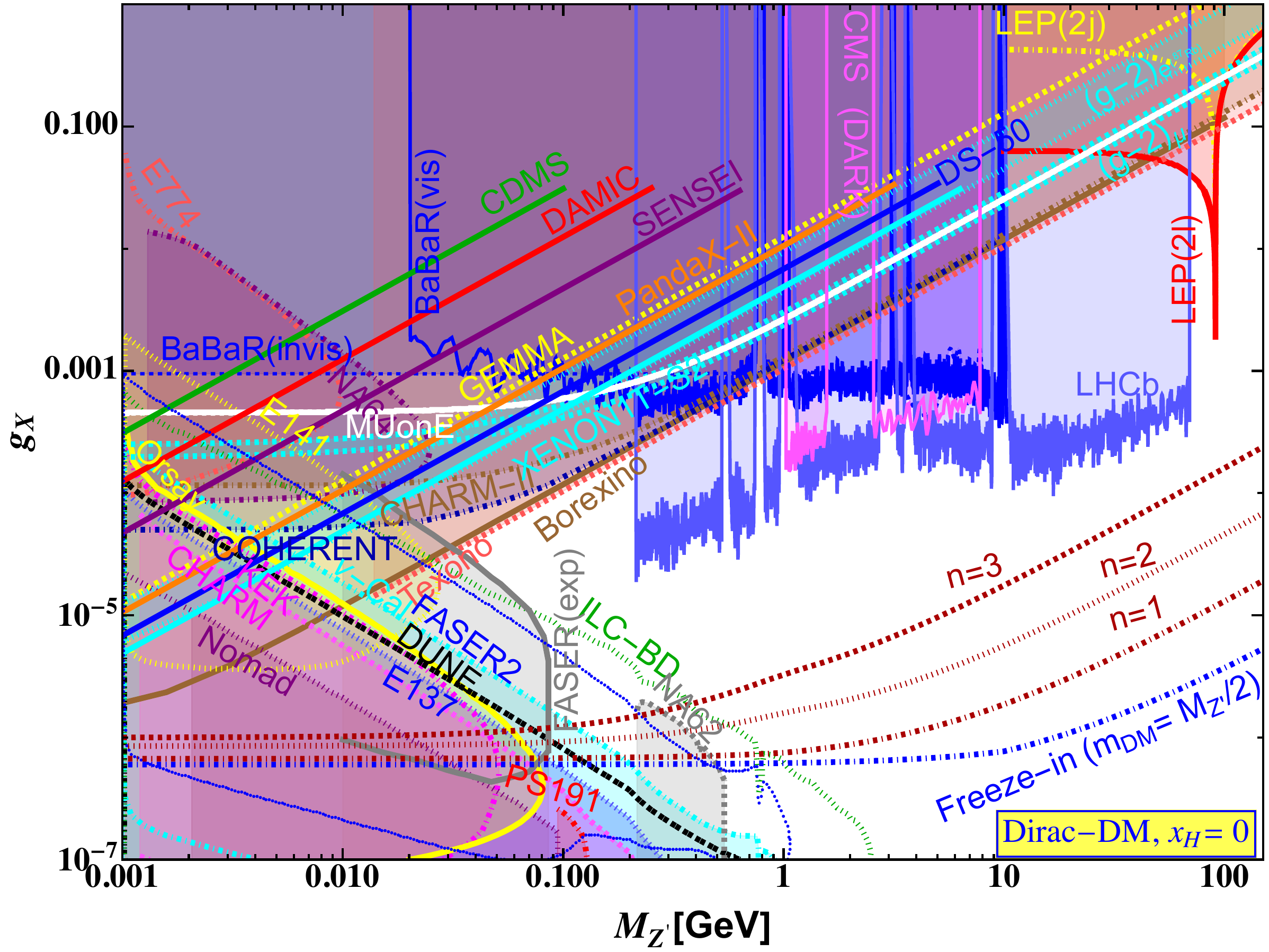}
\includegraphics[scale=0.203]{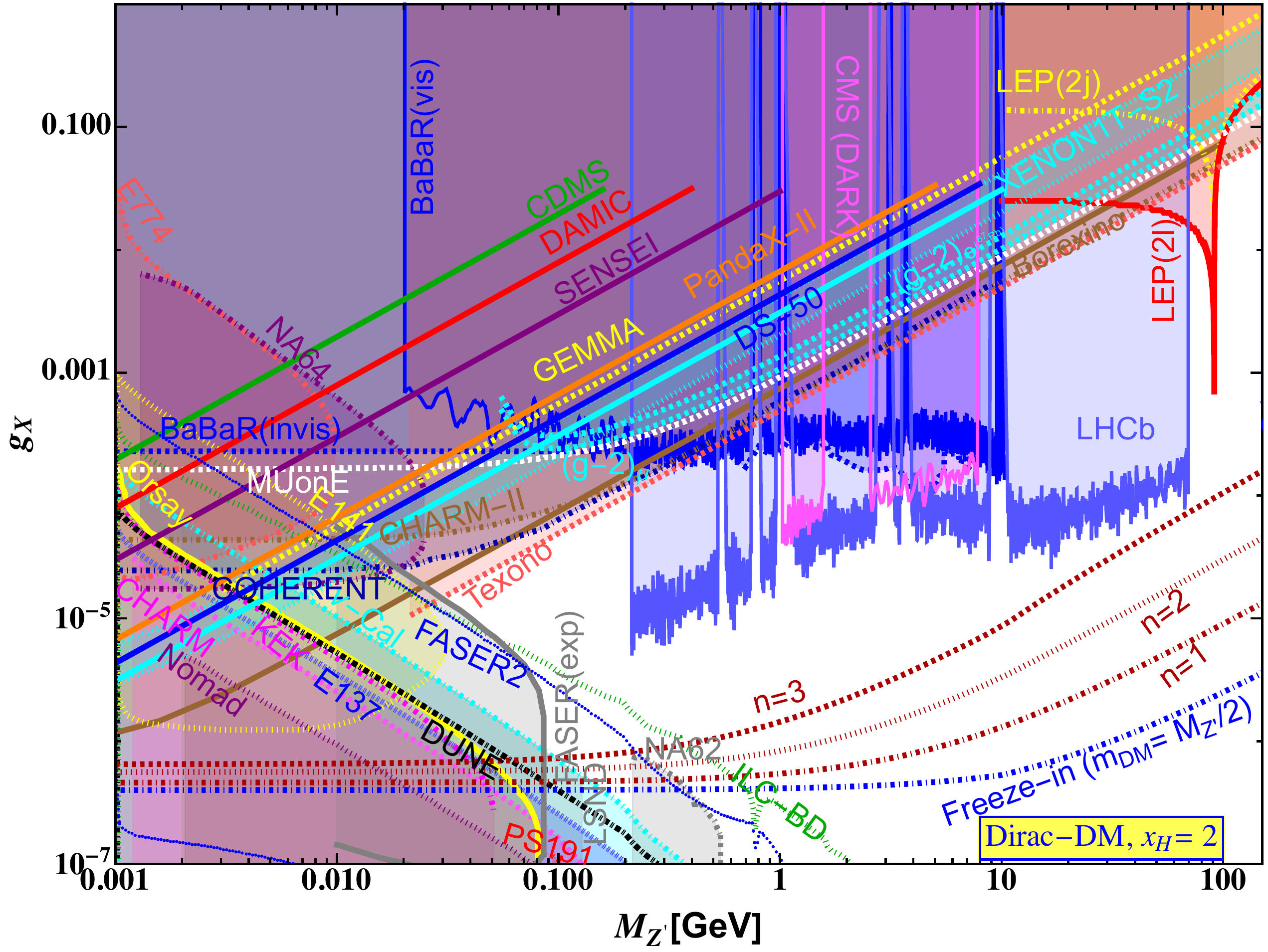}
\caption{Limits on $g_X-M_{Z^\prime}$ plane from DM-electron scattering process in $t$-channel mediated by $Z^\prime$ for $x_H=0$ $(2)$ considering $m_{\rm DM}=100$ MeV in left (right) panel for Majorana (upper), Scalar (middle) and Dirac (lower panel) DM candidates. Shaded regions are ruled out by existing experimental data whereas prospective bounds could appear from MUonE, DUNE, FASER2 and ILC-BD experiments. We add constraints obtained from Freeze-in mechanism considering $m_{\rm DM}=M_{Z^\prime}/2$ including modified cosmology.}
\label{lim2}
\end{figure}

We summarize the results in Figs.~\ref{lim1}-\ref{lim2}, where we show the bounds on $g_X-M_{Z^\prime}$ plane, for Majorana, scalar and Dirac DM from electron-DM scattering and freeze-in scenarios, respectively. In Fig.~\ref{lim1} we show the constraints for $x_H=-2$, the $U(1)_R$ case, where in case of electron-DM scattering we chose $m_{\rm DM}=10$(100) MeV shown in the left (right) panel for Majorana (upper), Scalar (middle) and Dirac DM (lower panel). We chose $x_H=-2$ because for this charge left handed fermions do not directly interact with $Z^\prime$ whereas right handed fermions do. As a result left handed neutrinos do not interact with $Z^\prime$ forbidding the invisible decay of $Z^\prime$. On the other hand we do the same for the $x_H=0$, the B$-$L scenario where left and right handed fermions interact with $Z^\prime$ in the same way. In addition to that we consider another $U(1)_X$ charge where $x_H=2$, where all the left and right handed fermions interact differently with $Z^\prime$. The corresponding bounds are shown in Fig.~\ref{lim2} for $x_H=0$ (2) in the left (right) panel for the Majorana(upper), Scalar(middle) and Dirac DM(lower panel)
considering $m_{\rm DM}=100$ MeV to study electron-DM scattering. In these analyses we considered the $U(1)_X$ charge of the Dirac DM to be $Q_\chi=100$. To estimate the constraints on $g_X-M_{Z^\prime}$ plane for different $x_H$ from electron-DM scattering and freeze-in scenarios we apply the perturbativity constraints $g_{X} (x_{H, \Phi}; Q_\chi) < \sqrt{4\pi}$. We have already mentioned that in this analysis we chose $x_\Phi=1$. Finally we compare our results with the constraints obtained from different scattering and beam-dump experiments. 

For the DM-electron scattering, we consider existing bounds from Super-CDMS~\cite{SuperCDMS:2020ymb}, DAMIC~\cite{DAMIC:2019dcn}, SENSEI~\cite{SENSEI:2020dpa}, PandaX-II~\cite{PandaX-II:2021nsg}, DarkSide-50 (DS-50)~\cite{DarkSide:2022knj} and XENON1T-S2~\cite{XENON:2019gfn}. For a given DM-type, we choose two benchmark values of DM mass: 10 MeV and 100 MeV. This is motivated by the fact that scattering experiments such as CDMS, DAMIC, and SENSEI impose the most stringent bounds for 10 MeV DM, while PandaX-II, DarkSide-50, and XENON1T-S2 offer greater sensitivity to DM with a mass of 100 MeV. This is illustrated in Fig.~\ref{fig:boundDD}. Therefore, in the $g_X-\Mzp$ plane, these benchmark DM masses shall provide the strongest constraints. For example, in the left (right) panel of  Fig.~\ref{lim1}, we consider $\mdm=10 (100)$ MeV, and bounds from Super-CDMS, DAMIC, SENSEI, PandaX-II, DarkSide-50 (DS-50) and XENON1T-S2 are shown via green, red, purple, orange, blue and cyan dashed diagonal straight lines respectively. In the right panel we show the same, but for a DM of mass 100 MeV. We find, irrespective of the DM-spin and mass, XENON1T-S2 provides most stringent bound for $x_H=-2$. In case of Majorana DM, XENON1T-S2 provide strongest limit on the general $U(1)_X$ coupling as $ 2.4 \times 10^{-3}\lesssim g_X \lesssim 4.3\times 10^{-3}$ with 0.012 GeV $\lesssim \Mzp \lesssim$ 0.0203 GeV for $\mdm=10$ MeV and $1.05 \times 10^{-3} \lesssim g_X\lesssim 1.8\times 10^{-3}$ with 0.02 GeV $\lesssim\Mzp\lesssim$ 0.033 GeV for $\mdm=100$ MeV. These limits are approximately same for the scalar case as well. Bounds below and above this range of $M_{Z^\prime}$ are ruled out by the existing bounds from NA64 \cite{NA64:2019auh} and the constraints obtained from the visible decay of $Z^\prime$ in BaBaR \cite{BaBar:2014zli,BaBar:2017tiz} experiment. On the other hand for Dirac DM because of the larger $U(1)_X$ charge, we see that the limits from different experiments have different end-points. Also, in this case, corresponding bounds on $g_X$ are stronger compared to the other two cases. We find that anything above $2.7\times 10^{-4}\lesssim g_X\lesssim 1.3\times 10^{-3}$ and $1.25 \times 10^{-4} \lesssim g_X\lesssim 10^{-3}$ for $\mdm=10$ MeV and 100 MeV within the range 0.01 GeV $\lesssim\Mzp\lesssim$ 0.045 GeV and 0.018 GeV $\lesssim\Mzp\lesssim$ 0.140 GeV, respectively are ruled out by XENON1T-S2 data. Bounds below and above this range of $M_{Z^\prime}$ are ruled out by E141 \cite{Riordan:1987aw} and visible decay of $Z^\prime$ in BaBaR \cite{BaBar:2014zli,BaBar:2017tiz}, respectively. In case of Dirac-DM scenario we find that the strongest bound obtained from XENON1T-S2 is slightly stronger than the limits obtained by the electron $g-2$ limits \cite{Asai:2023mzl}. We compare our results for the electron-DM scattering with the bounds estimated form the dark photon searches in LHCb \cite{Foguel:2022unm,Baruch:2022esd} and CMS experiments \cite{CMS:2023slr} from \cite{Asai:2023mzl}. These experiments strongly constrain  0.2 GeV $\lesssim M_{Z^\prime} \lesssim$ 70 GeV where limits on the $U(1)_X$ coupling vary between $10^{-5} \lesssim g_X \lesssim 10^{-4}$. We also find that constraints from electron $(g-2)$ \cite{Asai:2023mzl} can be slightly weaker than (comparable with) the limits estimated from the DM-electron scattering process in DS-50 (XENON1T-S2) experiment in case of Dirac-DM scenario for $m_{\rm DM}=100$ MeV whereas estimated bounds from other experiments are weaker than the limits from electron $(g-2)$. In case of Majorana and Scalar-DM scenarios, limits from DM-electron scattering is weaker than the bounds from electron $(g-2)$ for $m_{\rm DM}=10$ MeV and 100 MeV. Same results can be obtained for the Dirac-DM case for $m_{\rm DM}=10$ MeV. 

If we look at the scenarios for $x_H=0$ and $2$ in Fig.~\ref{lim2}, we find that limits obtained from invisible decay of $Z^\prime$ involving light neutrinos in BaBaR \cite{BaBar:2014zli,BaBar:2017tiz} rule out the limits obtained from electon-DM scattering. In this case we show the bounds estimated considering neutrino-electron scattering for TEXONO~\cite{TEXONO:2009knm,TEXONO:2002pra,TEXONO:2006xds}, BOREXINO~\cite{Borexino:2000uvj,Bellini:2011rx,Cleveland:1998nv,Lande:2003ex,Borexino:2007kvk,Borexino:2008gab} from \cite{Asai:2023mzl} due to complementarity and notice that these existing bounds rules out the limits obtained from the electron-DM scattering. This happens because for these charges light neutrinos directly interact with the $Z^\prime$ which are not possible in case of $x_H=-2$ as the $U(1)_X$ charge of the SM lepton doublet vanishes for $x_\Phi=1$. In the same line following \cite{Asai:2023mzl}
we show the limits on $g_X-M_{Z^\prime}$ plane for $x_H=0$ and 2 studying the muon neutrino, anti-neutrino scattering with electron from CHARM-II \cite{CHARM-II:1991ydz,CHARM-II:1989nic,CHARM-II:1993phx,CHARM-II:1994dzw} experiment. In this context we also compare the bounds obtained from neutrino-nucleon scattering from COHERENT \cite{COHERENT:2018imc,COHERENT:2018imc,Cadeddu:2017etk,Cadeddu:2020nbr,COHERENT:2020iec,COHERENT:2020ybo} experiment followed by considering the limits obtained from the neutrino magnetic moment study in GEMMA experiment \cite{Beda:2009kx,Lindner:2018kjo} from \cite{Asai:2023mzl}. To show complementarity we present the bounds obtained from the dark photon searches at LHCb and CMS experiments from \cite{Asai:2023mzl} for $x_H=0$ and 2. Limits obtained from DM-electron scattering for these charges are comparable with the electron $(g-2)$ limits for the Dirac-DM scenario for $m_{\rm DM}=10$ MeV and 100 MeV, however, it is stronger than the limits obtained from muon $(g-2)$ \cite{Asai:2023mzl}. Nature of the bounds for Majorana and scalar-DM scenarios are same as $x_H=-2$. However, it has been already pointed out that strong constraints from TEXONO and BOREXINO are also stronger than the $(g-2) $bounds. For $M_{Z^\prime} \lesssim 0.04$ GeV beam dump experiments provide stronger limits. 

Finally we show the bounds obtained from the dilepton and dijet searches in the LEP experiment \cite{ALEPH:2006bhb,ALEPH:1997gvm,LEPWorkingGroupforHiggsbosonsearches:2003ing,ALEPH:2005ab} following \cite{Asai:2023mzl,KA:2023dyz}, with sharp resonance at the $Z-$pole where $g_X$ could reach around $0.001$. In the context of comparing scattering scenarios we mention that there could be a future muon-electron scattering experiment called MUonE \cite{Asai:2023mzl} which could also provide stronger limit for $x_H=-2$ and hence can probe the strongest parameter region obtained from electron-DM scattering in XENON1T-S2 experiment in case of Dirac-DM scenario. The limits obtained from XENON1T-S2 for Dirac DM scenario could provide stonger sensitivity than MUonE experiment for 0.01 GeV $\lesssim M_{Z^\prime} \lesssim 0.025$ GeV which could be tested in future. However, at the current stage bounds obtained from Majorana and scalar DM scenarios will provide weaker constraints than the prospective MUonE experiment as shown in Fig.~\ref{lim1}. In case of $x_H=0$ and 2 we find from Fig.~\ref{lim2} that future constraints from MUonE experiment could be weaker than the existing limits from TEXONO, BOREXINO, COHERENT experiments, respectively. 

The freeze-in lines for each DM kind are shown in Fig.~\ref{lim1} for $x_H=-2$ and in Fig.~\ref{lim2} for $x_H=0$ (left panel) and $2$ (right panel) via the blue and brown dashed curves, corresponding to freeze-in production during radiation domination and during the epoch of modified cosmology, respectively. In all cases we consider $\mdm=2\,\Mzp$, such that DM production via $s$-channel $Z'$-mediated scattering dominates. As elaborated in Sec.~\ref{sec:freeze-in}, a modified cosmological background can improve the DM-SM coupling by orders of magnitude. This can be prominently seen for $\Mzp\gtrsim 0.1$ GeV, while for lighter $\Mzp$ the DM abundance becomes insensitive to the choice of the mediator mass. Following the prescription in subsection~\ref{sec:mod-cosmo}, a larger $n$ requires larger $g_X$ to produce the right DM abundance. As an instance, we see, for $n=3$ with $x_H=-2$, one needs $g_X\simeq 4\times 10^{-4}$ for $\Mzp=10$ GeV, while in case of standard cosmological background, $g_X\simeq 2\times 10^{-5}$ is required. For different $x_H$, it is not possible to go beyond $n=3$, as in that case, $g_X\gtrsim 0.1$ is needed to satisfy the observed DM abundance. For such a large coupling, the out of equilibrium production of DM becomes a question, and therefore freeze-in remains valid no more. In case of Majorana and scalar DM, we see, bounds form LHCb~\cite{Foguel:2022unm,Baruch:2022esd}, BaBaR \cite{BaBar:2014zli,BaBar:2017tiz}, CMS~\cite{CMS:2023slr} and LEP-II searches are already probing $n>1$ scenario for different choices of $x_H$ when $\Mzp\gtrsim 1$ GeV. For Dirac DM, however, $n=3$ remains still below bounds obtained from LHCb search. This is attributed to the large Dirac DM charge that in turn demands smaller $g_X$ to ensure that the DM does not overclose the Universe. For lighter $\Mzp<1$ GeV, electron/positron beam dump experiments, for example, Orsay~\cite{Davier:1989wz}, NA64~\cite{NA64:2019auh}, KEK~\cite{Beer:1986qr}, E141~\cite{Riordan:1987aw}, E137~\cite{Bjorken:1988as}, E774~\cite{Bross:1989mp}, are providing strong bounds on $g_X$ irrespective of the nature of the DM and for both modified $(n>0)$ and unmodified $(n=0)$ cosmological scenarios. Even after substantial improvement in the DM-SM coupling because of alternative cosmological scenario prior to BBN, the freeze-in lines are well beyond the reach of direct search limits in all cases. We also find that proton beam dump experiments involving Nomad~\cite{NOMAD:2001xxt}, CHARM~\cite{CHARM:1985anb}, $\nu-$cal~\cite{Blumlein:2011mv,Blumlein:2013cua}, NA62 \cite{NA62:2023qyn} and FASER \cite{FASER:2023tle} could probe the freeze-in lines for $M_{Z^\prime} < 1$ GeV. In addition to that, we compare our results with prospective bounds obtained from proton beam dump in FASER2, DUNE experiments and electron beam dump scenario at ILC (ILC-BD) experiment \cite{Asai:2022zxw} which could probe the freeze-in scenarios in future. 
\section{Conclusions} 
\label{sec:concl}
We consider a general $U(1)_X$ extension of the SM where three generations of RHNs are introduced to generate the neutrino mass through seesaw mechanism after spontaneous breaking of $U(1)_X$. On cancellation of gauge and mixed gauge-gravity anomalies we find, the $U(1)_X$ charge assignments of the left and right handed SM fermions are different. As a result, they interact differently with beyond the SM neutral gauge boson $Z^\prime$ of the model. We consider three different charges $x_H=-2,$ 0 and $2$, while fixing $x_\Phi=1$. In the first case, $U(1)_X$ charges of the left handed SM fermions vanish manifesting the $U(1)_R$ scenario, in the second case, left and right handed SM fermions interact with $Z^\prime$ in the same way manifesting B$-$L scenario; the third case is a chiral one where left and right handed SM fermions interact differently with $Z^\prime$. To explore the phenomenological implications of this property, we consider electron-DM scattering processes mediated by $Z^\prime$ for Majorana, scalar and Dirac type DM, to estimate constraints on $g_X-M_{Z^\prime}$ plane and compare with existing bounds from several electron-DM scattering experiments, considering two benchmark choices of DM masses [cf. Fig.~\ref{fig:boundDD}]. In order to check the consistency between theoretical estimation and experimental predictions, we have also estimated events rates under different experimental set-up for electron-DM scattering [cf. Fig.~\ref{fig:event_rate}]. Finally, we find, in case of $x_H=-2$ electron-DM scattering provides strong bounds for different DM models. Among them, strongest bound is obtained from the XENON1T-S2 experiment. However, in case of $x_H=0$ and 2, we find that neutrino-electron and neutrino-nucleon scattering experiments provide strong constrains on $g_X-M_{Z^\prime}$ plane which are stronger than the limits estimated from electron-DM scattering. Therefore depending on the electron-DM scattering experiment, strongest limits could be provided in $U(1)_R$ scenario while the parameter space is compared to the scattering and beam-dump experiments approximately for the range 0.01 GeV $ \lesssim M_{Z^\prime} \lesssim 1$ GeV for Dirac-DM. However, this window of $M_{Z^\prime}$ is narrow in case of Majorana and Scalar type DM models. 

In the same framework, we also study freeze-in production via $\rm{SM~SM} \to \rm{DM~ DM}$ scattering (mediated by $Z'$) in a radiation dominated and in a modified cosmological background, reproducing observed DM relic abundance forbidding the decay of $Z^\prime$ into DM candidates [cf. Fig.~\ref{fig:yield}]. In the modified cosmological scenario, we consider the Universe expands faster compared to the standard radiation domination prior to the onset of big bang nucleosynthesis (BBN). As before, we estimate bounds on $g_X-M_{Z^\prime}$ plane for $x_H=-2$, 0 and $2$ taking different DM candidates into account. We see, existing bounds from electron and proton beam dump experiments rules out freeze-in predictions for $M_{Z^\prime} \lesssim 0.1$ GeV, whereas depending on $x_H$ and nature of the DM candidates these bounds become weaker for $M_{Z^\prime} \gtrsim 10$ GeV. In case of Majorana and scalar DM candidates, bound from freeze-in production in modified cosmological background is ruled out by existing LHCb constraints, while the bounds obtained from freeze-in production during radiation domination stay beyond the current LHCb reach. However, in case of a Dirac DM, the estimated bounds from freeze-in are always strong for 0.8 GeV $\lesssim M_{Z^\prime} \lesssim 150$ GeV and region for $M_{Z^\prime} \lesssim 0.8$ GeV is ruled out by existing limits form beam dump experiments. We thus infer, present bounds from beam dump, LHCb, CMS, BaBaR and LEP can potentially put bound on pre-BBN modification to standard cosmology, thereby constraining different models of modified gravity and cosmology. However, freeze-in couplings required to produce the observed DM abundance remains well beyond the reach of DM-electron scattering experiments [cf. Fig.~\ref{lim1} and \ref{lim2}]. Proton beam dump experiments FASER2, DUNE and electron beam dump experiment in ILC could probe parameter regions on $g_X-M_{Z^\prime}$ plane for different general $U(1)_X$ charges, providing a complementarity in future.\\  

\noindent
{\textbf{Acknowledgments.--}} The work of S.M. is supported by KIAS Individual Grants (PG086002) at Korea Institute for Advanced Study.
\begin{widetext}
\appendix
\section{Kinematics and scattering amplitudes}
\label{sec:app-kin}
We consider the DM$-e$ elastic scattering process DM\,$e^- \to e^-$\,DM in the $t-$channel mediated by $Z^\prime$. We define the four momenta in the laboratory frame as
\begin{align}
& p_1 = \left(\Edm,\,\vec{p}_1\right),~~
p_2 = \left(m_e,\,\vec{0}\right),~~
p_3 = \left(\Edm',\,\vec{p}_3\right)
p_4 = \left(E_4,\,\vec{p}_4\right)
\end{align}
From 4-momenta conservation we further obtain
\begin{align}
& \Edm'=\Edm+m_e-E_4 \text{ with } E_4=m_e+E_r,
\\&
p_1\cdot p_2=p_3\cdot p_4=\Edm\,m_e,~~
p_3\cdot p_2=p_4\cdot p_1=m_e\,\left(\Edm+m_e-E_4\right)
\\&
p_1\cdot p_3=\mdm^2-m_e^2+m_e\,E_4,~~
p_2\cdot p_4=m_e\,E_4, ~~
q^2=\left(p_4-p_2\right)^2=-2m_e\,\Erec\,.
\end{align}
The differential cross-section in the laboratory frame reads 
\begin{align}\label{eq:dsdEr}
& \frac{d\sigma}{d\Erec}
=\frac{1}{32\,|\vec{p}_1|^2\,m_e}\,\left|\overline{\mathcal{M}}\right|^2_{\text{DM}e\to e\text{DM}}\,,
\end{align}
where
\begin{align}
& \left|\overline{\mathcal{M}}\right|^2_{\chi e\to e\chi}=\frac{g_X^4\, Q_\chi^2\,m_e}{\left(2\,\Erec\,m_e+\Mzp^2\right)^2+\Mzp^2\,\Gamma_{Z'}^2}\Big[
\left(5x_H^2+12x_H x_\Phi+8 x_\Phi^2\right)
\Big\{\Edm^2 m_e+m_e (\Edm-\Erec)^2-
\nonumber\\&
m_\chi^2 (\Erec+m_e)\Big\}
-4\,m_e\,(x_H+x_\Phi)\,(x_H+2 x_\Phi) 
\left(\Erec m_e+m_\chi^2\right)
+8 m_e m_\chi^2 (x_H+x_\Phi) (x_H+2 x_\Phi)\Big]\,,
\end{align}
is for Dirac DM,
\begin{align}
&
\left|\overline{\mathcal{M}}\right|^2_{Ne\to eN}=\frac{g_X^4\,x_\Phi^2\,m_e}{\left(2\,\Erec\,m_e+\Mzp^2\right)^2+\Mzp^2\,\Gamma_{Z'}^2}
\Big[\left(5 x_H^2+12 x_H x_\Phi+8 x_\Phi^2\right)\times 
\Big\{\Edm^2 m_e+m_e (\Edm-\Erec)^2+
\nonumber\\&
M_N^2 (\Erec+m_e)\Big\}-4 m_e (x_H+x_\Phi) (x_H+2 x_\Phi) \left(\Erec m_e+M_N^2\right)
-8 m_e M_N^2 (x_H+x_\Phi) (x_H+2 x_\Phi)\Big]\,,
\end{align}
is for Majorana DM and
\begin{align}   
&
\left|\overline{\mathcal{M}}\right|^2_{\Phi e\to e\Phi}=\frac{g_X^4\,x_\Phi^2\,m_e}{\left(2\,\Erec\,m_e+\Mzp^2\right)^2+\Mzp^2\,\Gamma_{Z'}^2}\Big[
x_H^2\Big\{20 \Edm\,m_e\,(\Edm-\Erec)-\Erec m_e^2
-2 m_\Phi^2 (5 \Erec+m_e)\Big\}+ 
\nonumber\\&
24 x_H x_\Phi \Big\{2 \Edm m_e (\Edm-\Erec)-\Erec m_\Phi^2\Big\}
+16 x_\Phi^2 \Big\{2 \Edm m_e (\Edm-\Erec)-\Erec m_\Phi^2\Big\}\Big]\,,
\end{align}
in case of scalar DM. 
\section{Scattering cross-section}
\label{sec:app-scattering}
Here we report expressions for  the scattering cross-sections in the center of mass frame of a pair of massless SM fermions $(f)$ going into a pair of DM particles as $f \bar{f} \to$ DM DM, mediated by $s$-channel $Z^\prime$ where $f=\ell$ (charged lepton), $\nu$ (neutrino), $u$ (up quark) and $d$ (down quark), respectively. For charged lepton initial states we have
\begin{align}
& \sigma(s)_{\ell^+\ell^-\to\text{DM}\text{DM}}\simeq \frac{g_X^4}{96\pi\,\left[(s-\Mzp^2)^2+\Gamma_{Z'}^2\,\Mzp^2\right]}\,\sqrt{1-\frac{4\,\mdm^2}{s}}
\begin{cases}
Q_\chi^2\,\left(s+2\,\mdm^2\right)\,\left(5\,x_H^2+12\,x_H\,x_\Phi+8\,x_\Phi^2\right) & \text{for Dirac}
\\[10pt]
x_\Phi^2\,\frac{\left(s-4\,\mdm^2\right)\,\left(5\,x_H^2+12\,x_H\,x_\Phi+8\,x_\Phi^2\right)}{2} & \text{for Majorana}
\\[10pt]
x_\Phi^2\,\frac{\left(s-4\,\mdm^2\right)\,\left(5\,x_H^2+12\,x_H\,x_\Phi+8\,x_\Phi^2\right)}{4} & \text{for Scalar}\,.
\end{cases}
\end{align}
For light neutrino initial states
\begin{align}
& \sigma(s)_{\nu\nu\to\text{DM}\text{DM}}\simeq \frac{g_X^4}{48\pi\,\left[(s-\Mzp^2)^2+\Gamma_{Z'}^2\,\Mzp^2\right]}\,\sqrt{1-\frac{4\,\mdm^2}{s}}
\begin{cases}
Q_\chi^2\,\left(s+2\,\mdm^2\right)\,\left(x_H+2\,x_\Phi\right)^2 &\text{for Dirac}
\\[10pt]
x_\Phi^2\,\frac{\left(s-4\,\mdm^2\right)\,\left(x_H+2\,x_\Phi\right)^2}{4} & \text{for Majorana}
\\[10pt]
x_\Phi^2\,\frac{\left(s-4\,\mdm^2\right)\,\left(x_H+2\,x_\Phi\right)^2}{4} & \text{for Scalar}\,.
\end{cases}
\end{align}
For up-quark initial states,
\begin{align}
& \sigma(s)_{uu\to\text{DM}\text{DM}}\simeq \frac{g_X^4}{2592\pi\,\left[(s-\Mzp^2)^2+\Gamma_{Z'}^2\,\Mzp^2\right]}\,\sqrt{1-\frac{4\,\mdm^2}{s}}
\begin{cases}
Q_\chi^2\,\left(s+2\,\mdm^2\right)\,\left(17\,x_H^2+20\,x_H\,x_\Phi+8\,x_\Phi^2\right) &\text{for Dirac}
\\[10pt]
x_\Phi^2\,\frac{\left(s-4\,\mdm^2\right)\,\left(17\,x_H^2+20\,x_H\,x_\Phi+8\,x_\Phi^2\right)}{2} & \text{for Majorana}
\\[10pt]
x_\Phi^2\,\frac{\left(s-4\,\mdm^2\right)\,\left(17\,x_H^2+20\,x_H\,x_\Phi+8\,x_\Phi^2\right)}{4} & \text{for Scalar}\,.
\end{cases}
\end{align}
For down-quark initial states,
\begin{align}
& \sigma(s)_{dd\to\text{DM}\text{DM}}\simeq \frac{g_X^4}{2592\pi\,\left[(s-\Mzp^2)^2+\Gamma_{Z'}^2\,\Mzp^2\right]}\,\sqrt{1-\frac{4\,\mdm^2}{s}}
\begin{cases}
Q_\chi^2\,\left(s+2\,\mdm^2\right)\,\left(5\,x_H^2-4\,x_H\,x_\Phi+8\,x_\Phi^2\right) &\text{for Dirac}
\\[10pt]
x_\Phi^2\,\frac{\left(s-4\,\mdm^2\right)\,\left(5\,x_H^2-4\,x_H\,x_\Phi+8\,x_\Phi^2\right)}{2} & \text{for Majorana}
\\[10pt]
x_\Phi^2\,\frac{\left(s-4\,\mdm^2\right)\,\left(5\,x_H^2-4\,x_H\,x_\Phi+8\,x_\Phi^2\right)}{4} & \text{for Scalar}\,.
\end{cases}
\end{align}
\section{Reaction densities}
\label{sec:app-reacden}
The reaction density for to $1\to2$ decay process is given by
\begin{align}
\gamma_{12} &=\int\,\prod_{i=1}^3\,(2\,\pi)^4\,\delta^{(4)}\,\left(p_a-p_1-p_2\right)\,f_a^\text{eq}\,\left|\mathcal{M}\right|^2_{a\to1,2}=\frac{g_a}{2\,\pi^2}\,m_a^2\,\Gamma_{a\to1,2}\,T\,K_1\left(\frac{m_a}{T}\right)\,. 
\end{align}

The reaction density corresponding to 2-to-2 processes reads
\begin{align}
& \gamma_{22}=\int\prod_{i=1}^4 d\Pi_i \left(2\pi\right)^4 \delta^{(4)}\biggl(p_a+p_b-p_1-p_2\biggr)
f_a{^\text{eq}}f_b{^\text{eq}}\left|\mathcal{M}_{a,b\to1,2}\right|^2
\nonumber\\&
=\frac{T}{32\pi^4}\,g_a g_b\,\int_{s_\text{min}}^\infty ds\,\frac{\biggl[\bigl(s-m_a^2-m_b^2\bigr)^2-4m_a^2 m_b^2\biggr]}{\sqrt{s}}
\sigma\left(s\right)_{a,b\to1,2}\,K_1\left(\frac{\sqrt{s}}{T}\right)\label{eq:gam-ann}\,,
 \end{align}    
with $a,b(1,2)$ as the incoming (outgoing) states and $g_{a,b}$ are corresponding degrees of freedom. Here $f_i{^\text{eq}}\approx\exp^{-E_i/T}$ is the Maxwell-Boltzmann distribution. The Lorentz invariant 2-body phase space is denoted by: $d\Pi_i=\frac{d^3p_i}{\left(2\pi\right)^3 2E_i}$. The amplitude squared (summed over final and averaged over initial states) is denoted by $\left|\mathcal{M}_{a,b\to1,2}\right|^2$ for a particular 2-to-2 scattering process. The lower limit of the integration over $s$ is $s_\text{min}=\text{max}\left[\left(m_a+m_b\right)^2,\left(m_1+m_2\right)^2\right]$.
\end{widetext}
\bibliographystyle{utphys}
\bibliography{references}
\end{document}